\documentclass[a4paper,fleqn,twocolumn,margin=1in]{aastex61}
\pdfoutput=1
\usepackage{float}
\usepackage{graphicx}
\usepackage{ae,aecompl} 
\usepackage{enumitem} 
\usepackage{amssymb} 
\usepackage{amsmath} 
\usepackage[bookmarks=false]{hyperref} 

\def\apjl{\rm ApJL} 
\def\apjs{\rm ApJS}

\def\aap{\rm AAP}

\def\gax{\mathrel{\raise.3ex\hbox{$>$}\mkern-14mu\lower0.6ex\hbox{$\sim$}}} 
\def\lax{\mathrel{\raise.3ex\hbox{$<$}\mkern-14mu\lower0.6ex\hbox{$\sim$}}} 
\def\gtorder{\mathrel{\raise.3ex\hbox{$>$}\mkern-14mu 
             \lower0.6ex\hbox{$\sim$}}} 
\def\ltorder{\mathrel{\raise.3ex\hbox{$<$}\mkern-14mu 
             \lower0.6ex\hbox{$\sim$}}}

\def\plx{\varpi}
\def\varpigaia{\varpi_{Gaia}}
\def\hatvarpigaia{\hat{\varpi}_{Gaia}}

\def\muhz{\mu\mathrm{Hz}}
\def\numax{\nu_{\mathrm{max}}}
\def\hatnumax{\hat{\nu}_{\mathrm{max}}}
\def\dnu{\Delta \nu}
\def\hatdnu{\hat{\Delta \nu}}

\def\teff{T_{\mathrm{eff}}}
\def\hatteff{\hat{T}_{\mathrm{eff}}}

\def\teffsun{T_{\mathrm{eff,} \odot}}

\def\rsun{R_{\odot}}
\def\deg{^{\circ}}

\def\numaxsun{\nu_{\mathrm{max,} \odot}}
\def\dnusun{\Delta\nu_{\odot}}

\def\hatinvrscal{\hat{R}^{-1}_{\mathrm{seis}}}
\def\hatinvrscali{\hat{R}^{-1}_{\mathrm{seis},i}}

\def\hatinvrgaia{\hat{R}^{-1}_{Gaia}}
\def\hatinvrgaiai{\hat{R}^{-1}_{Gaia,i}}
\def\rscal{R_{\mathrm{seis}}}

\def\varpiast{\varpi_{\mathrm{seis}}}
\def\hatvarpiast{\hat{\varpi}_{\mathrm{seis}}}
\def\muas{\mu \mathrm{as}}
\def\mas{\mathrm{mas}}
\def\nueff{\nu_{\mathrm{eff}}}
\def\hatnueff{\hat{\nu}_{\mathrm{eff}}}
\def\ks{K_{\mathrm{s}}}
\begin{document}
\title{Confirmation of the {\it Gaia} DR2 parallax zero-point offset 
  using asteroseismology and spectroscopy in the
  {\it Kepler} field}
\author{Joel C. Zinn}
\affiliation{Department of Astronomy, The Ohio State University, 140 West
  18th Avenue, Columbus OH 43210, USA}
\author{Marc H. Pinsonneault}
\affiliation{Department of Astronomy, The Ohio State University, 140 West
  18th Avenue, Columbus OH 43210, USA}
\author{Daniel Huber}
\affiliation{ Institute for Astronomy, University of Hawai`i, 2680 Woodlawn Drive, Honolulu, HI 96822, USA}
\affiliation{Sydney Institute for Astronomy (SIfA), School of Physics,
University of Sydney, NSW 2006, Australia}
\affiliation{Stellar Astrophysics Centre, Department of Physics and
Astronomy, Aarhus University, Ny Munkegade 120, DK-8000
Aarhus C, Denmark}
\affiliation{SETI Institute, 189 Bernardo Avenue, Mountain View, CA
  94043, USA}
\author{Dennis Stello}
\affiliation{School of Physics, University of New South Wales, Barker
  Street, Sydney, NSW 2052, Australia}
\affiliation{Sydney Institute for Astronomy (SIfA), School of Physics,
  University of Sydney, NSW 2006, Australia}
\affiliation{Stellar Astrophysics Centre, Department of Physics and
Astronomy, Aarhus University, Ny Munkegade 120, DK-8000
Aarhus C, Denmark}
\affiliation{Center of Excellence for Astrophysics in Three Dimensions (ASTRO-3D), Australia}
\correspondingauthor{Joel C. Zinn}
\email{zinn.44@osu.edu}

\accepted{3 May 2019}
\submitjournal{The Astrophysical Journal}  

\begin{abstract}
We present an independent confirmation of the zero-point offset of
{\it Gaia} Data Release 2 (DR2) parallaxes using asteroseismic data of
evolved stars in the {\it Kepler} field. Using well-characterized red
giant branch (RGB) stars from the APOKASC-2 catalogue we identify a {\it Gaia} astrometric pseudo-color ($\nueff$)- and
{\it Gaia} $G$-band magnitude-dependent zero-point offset of
  $\varpiast - \varpigaia = 52.8 \pm 2.4 {\rm\ (rand.)} \pm  8.6 {\rm\ (syst.)} - (150.7 \pm  22.7)(\nueff - 1.5) -  (4.21 \pm 0.77)(G - 12.2)  \muas$, in the sense that {\it Gaia} parallaxes are %STAT
  too small. The offset is found in high
  and low-extinction samples, as well as among both shell H-burning
  red giant stars and core He-burning red
  clump stars. We
  show that errors in the asteroseismic radius and temperature scales may be distinguished
  from errors in the {\it Gaia} parallax scale. We estimate
  systematic effects on the inferred global {\it Gaia} parallax offset,
  $c$, due to radius and temperature systematics, as well as choices
  in bolometric correction and the adopted form for {\it Gaia}
  parallax spatial correlations. Because of possible spatially-correlated parallax errors, as discussed by the {\it Gaia} team, our
  {\it Gaia} parallax offset model is specific to the \textit{Kepler} field, but
  broadly compatible with the magnitude- and color-dependent offset inferred
  by the {\it Gaia} team and several subsequent investigations using independent methods. 
\end{abstract}

\section{Introduction}
\label{sec:intro}
The recent release of {\it Gaia} astrometry as part of Data Release 2 \citep{gaia2016,gaia2018}
signals an unprecedented opportunity to test stellar astrophysics. In
particular, the parallaxes --- with typical formal precisions of
$0.03\mas$ for sources with $G > 15$ (\citealt{lindegren+2018};
hereafter L18) --- can be used to solve one of the biggest
and most challenging problems in stellar astrophysics, namely the
determination of distances. At small parallax, however, the results
become sensitive to systematic errors, and checks from alternative
techniques are valuable.  In this paper we use asteroseismic data to
test zero-point offsets in the {\it Gaia} parallaxes.

The first data release of {\it Gaia},
  using the Tycho-{\it Gaia} astrometric solution (TGAS)
  \citep{michalik_lindegren&hobbs2015,gaia2016} represented a
  significant advance over the earlier {\it Hipparcos} work \citep{vanleeuwen2007}.  However, the
  TGAS investigators did note the existence of both spatial
  correlations and a zero-point offset \citep{lindegren+2016}. Their
  work was confirmed by other investigators. For the closest objects, \cite{jao+2016} and \cite{stassun&torres2016b} found consistent offsets
of $\approx 0.2\mas$ in the sense that TGAS parallaxes were too small when compared to
trigonometric parallaxes
for 612 dwarfs with parallaxes greater than $10 \mas$, and 111 eclipsing
binaries with parallaxes mostly greater than $1 \mas$, respectively. Comparing these
results for relatively nearby stars to results from more distant
giants with parallaxes of less than $1 \mas$ derived from {\it Kepler} \citep{borucki+2010} data indicated the presence of a
fractional zero-point offset \citep{deridder+2016,davies+2017,huber+2017}. Indeed, at larger
distances than the {\it Kepler} giant samples,
\cite{sesar+2017} found RR Lyrae parallaxes to show no indications of
an offset with TGAS parallaxes, and neither did \cite{casertano+2017} among a sample of Cepheid parallaxes.

There were also follow-up tests of spatially correlated parallaxes
after the publication of TGAS. \cite{jao+2016} confirmed these spatial
correlations in pointing out parallax offsets between hemispheres. \cite{casertano+2017} later reported evidence for spatial correlations
in the parallax error below $10^{\circ}$. Using a larger sample, \cite{zinn+2017} mapped out the spatial correlation of the errors in the {\it
  Kepler} field below $10\deg$ using asteroseismic distances of $\sim
1400$ giants, which showed correlations that increased
at sub-degree scales.

Systematic errors in the {\it Gaia} parallaxes exist, as well. Indications
are that a zero-point error might best be explained by a
degeneracy in the astrometric solution between a global parallax shift
and a term describing a periodic variation of the spacecraft's basic
angle\footnote{The angle between the two fields of view of {\it
    Gaia} that allows an absolute measure of parallax. See \cite{gaia2016} for a review of the mission design.}with a
period of the spacecraft spin period (L18). A smaller contribution
might arise from smearing of the PSF in the across-scan direction
(L18). As part of the DR2 release, \cite{arenou+2018} inferred several
estimates of a zero-point offset by comparing the
{\it Gaia} DR2 parallaxes to parallaxes of dwarf
galaxies, classical pulsators, stars in spectroscopic surveys, and
open \& globular clusters (see their Table 1). The zero-point offset does
vary among these sources, from $10\muas$ to $100\muas$, which may
represent genuine variation as a function of position on the sky,
magnitude, or color, or
various systematic errors in the comparison parallaxes.

\begin{figure}
\includegraphics[width=0.5\textwidth]{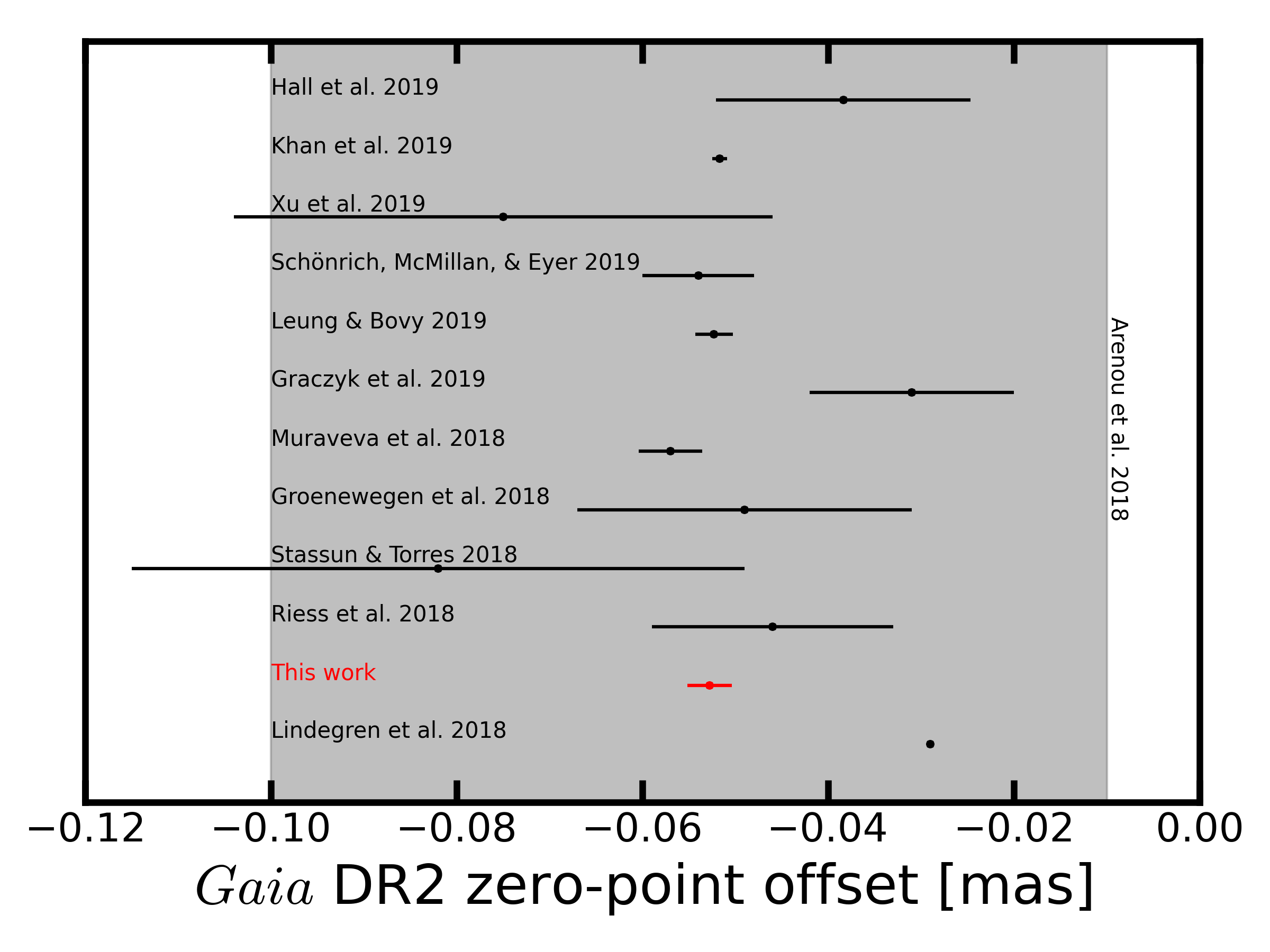} 
\caption{{\it Gaia} zero-point offsets from the literature, and their statistical uncertainties. An uncertainty was not given for the L18 result. The grey band indicates the range of the offset found by \protect\cite{arenou+2018}. See text for details.}
\label{fig:zps}
\end{figure}

Independent follow-up points to a similar magnitude for the parallax
zero-point systematic error. \cite{riess+2018} confirmed a global offset of $46 \pm 11 \muas$ for parallaxes
in {\it Gaia} by comparing {\it Gaia} parallaxes to those of a sample
of 50 Cepheids, whose distances can be precisely determined using a
period-luminosity relation. This particular sample is redder and
brighter than the
sample of quasars used in L18 to test the parallax systematics, and
may indeed have a genuinely different zero-point error due to trends
in parallax offsets with color and magnitude noted in
L18. Also using classical cepheids, \cite{groenewegen+2018a} determined a zero-point of $49\muas \pm 18\muas$, consistent with that of \cite{riess+2018}.  \cite{muraveva+2018a} similarly estimated a mean zero-point offset of $57\muas \pm 3.4\muas$ using three different RR Lyrae absolute magnitude relations\footnote{The uncertainty has been calculated from the standard deviation of the three methods used in \cite{muraveva+2018a}}. \cite{stassun+2018a} reported a global offset of $82 \pm
33\muas$ when comparing to a sample of 89 eclipsing binaries with
dynamical radii. Working with empirical eclipsing binary surface brightness--color relations, \cite{graczyk+2019} estimated a zero-point offset of $31\muas \pm 11\muas$. Other results confirm this picture: using a statistical approach based on the effect of parallax errors on tangential velocities, \cite{schonrich+2019a} determined a zero-point offset of $54\muas \pm 6.0\muas$; applying a machine learning distance classifier using APOGEE spectroscopy, \cite{leung+2019a} found a zero-point offset of $52.3\muas \pm 2.0\muas$; and \cite{xu+2019a} inferred an offset of $75\muas \pm 29\muas$ with very long baseline interferometry astrometry.  Most recently, \cite{khan+2019} and \cite{hall+2019} found offsets of $51.7\muas \pm 0.8\muas$ ($\pm \approx 10\muas$ when including spatially-correlated parallax errors) and $38.38^{+13.54}_{-13.83}\muas$, by computing asteroseismic parallaxes for {\it Kepler} RGB and RC stars, respectively. All these results are in a consistent direction, in the sense that {\it Gaia} DR2 parallaxes are too small, and, combined, yield a mean inferred offset of $53.6\muas$ and a variance-weighted mean of $51.9\muas$. We show these zero-point estimates from the literature in Figure~\ref{fig:zps}.

The {\it Gaia} team has quantified the parallax error budget in DR2
using almost 600,000 quasars from AllWISE \citep{secrest+2015}. They estimate
both a global zero-point error of $29 \muas$ (in the
sense that {\it Gaia} parallaxes are too small) and a spatial covariance
of the parallaxes, which have a typical angular scale of ten degrees
and an error of $10 \muas$, and which increases exponentially for smaller
scales. Crucially, this means that one cannot benefit from a
$\sqrt{N}$ reduction in random uncertainties of the parallax. Given these systematics, the {\it
  Gaia} team recommends adopting an irreducible systematic error on the
parallaxes of $\sim 0.1 \mas$ that takes into account both zero-point and
spatially-correlated errors. This recommended systematic error is large enough to
marginalize over much of the position-, color- and magnitude-dependent
nature of the systematics, and in that sense is likely larger than the
systematics particular to a specific data set and region of the sky.

Because of the large body of research performed in the {\it Kepler}
field, it is of great interest to quantify the particular systematic
errors among its giant population. Here, we quantify a zero-point offset
with a sample of nearly $3500$ giants with %STAT
precise asteroseismic distances in the {\it Kepler} field that also have {\it Gaia} parameters.

\section{Data}
\label{sec:data}
\subsection{The asteroseismic comparison sample}
Solar-like oscillations have been detected in thousands of evolved
stars by the
{\it CoRoT} and \textit{Kepler} missions
\citep{hekker+2009,de_ridder+2009,
  bedding+2010,stello+2013,yu+2018}. The overall properties of the
oscillation frequencies can be characterized by two global measurements:
the frequency of maximum power, $\numax$, and the large
frequency spacing, $\dnu$.  The observed frequency of maximum
power is related to both the surface gravity and $\teff$
\citep{kjeldsen&bedding1995,brown+1991}, while it can be demonstrated that the square of $\dnu$ is proportional to the mean density in the limiting case of
homology and large radial order, $n$ \citep{ulrich1986}.  We can therefore
solve for stellar mass and radius separately through the usage of
scaling relations, typically measured relative to the Sun, if we have
asteroseismic data and a robust effective temperature indicator.

Asteroseismic distance estimates are then possible because the
combination of radius and $\teff$ yields a luminosity. When
combined with an apparent magnitude, an appropriate set of bolometric
corrections, and an extinction, the distance can be
derived. Fortunately, all of these quantities are well-measured in the
{\it Kepler} field.

The basis of our data set is a sample of 6676 {\it Kepler} red giants with asteroseismic and
spectroscopic data taken from \citet{pinsonneault+2018}, hereafter APOKASC-2. The APOKASC-2 study
provides asteroseismic evolutionary state classification, masses,
radii, and extinction measures in the V-band. Of particular importance
is that the asteroseismic radii are verified to be on an absolute
scale by calibrating against fundamental data in star clusters, with typical
random uncertainties in radius of under 2\% and well-controlled %STAT
systematics. This means that our asteroseismic distances will also be
on a fundamental scale, which is ultimately tied to dynamical open
cluster masses.

As a complement to the asteroseismic information, there is uniform spectroscopic data from the
APOGEE survey of the Sloan Digital Sky Survey (SDSS) in the \textit{Kepler} field. Almost all
of the asteroseismic targets have photometry from 2MASS \citep{skrutskie+2006},
WISE \citep{wright+2010}, and $griz$ photometry from the {\it Kepler} Input Catalog
\citep{brown+2011} as corrected by \citet{pinsonneault+2012a} to be on the SDSS
system, and for which uncertainties are estimated to be $1\%$ in $g$
and $r$. The spectroscopic $\teff$ values from APOGEE that we use are calibrated to be in
agreement with the Infrared Flux Method (IRFM) photometric scale of
\citet{ghb09} for targets in low-extinction fields.  The
extinctions are well-studied in the \textit{Kepler} fields, because they can
be inferred by requiring consistency between photometric and
spectroscopic temperature estimates: extinction will redden photometry of an individual star, biasing its photometric temperature, so an extinction may be derived by adjusting the reddening until the photometric and spectroscopic temperatures agree \citep[see][for details on
  Bayesian fitting of the extinctions used in the APOKASC-2 catalogue]{rodrigues+2014}.

The masses and radii of shell H-burning (hereafter RGB) stars in the
APOKASC-2 catalogue are
computed using asteroseismic scaling relations, with reference values of $\numaxsun =
3076 \muhz$, $\dnusun = 135.146 \muhz$, and $\teffsun = 5772 K$. APOKASC-2 uses theoretically-motivated corrections to
the $\dnu$ scaling relation \citep[e.g.,][]{white+2011,sharma+2016}, which induce a differential change between the
radii and masses of RGB stars and core He-burning, or red clump (RC) stars. As a
result, it is important to analyze RGB and RC stars separately, as
there are known effects that can produce relative offsets between the
two populations \citep[e.g.,][]{miglio+2012}.  The relative radii of both are consistent with one
another in open clusters in the \textit{Kepler} field, but
differences at the few percent level could not be ruled out using
those samples alone. Our basic sample therefore consists of $3475$
RGB %STAT
stars and $2587$ RC stars which pass the {\it Gaia} DR2 selection cuts
as %STAT
described below. Unless otherwise noted, we only use first-ascent red
giants in our main analysis because the APOKASC-2 radii are directly calibrated
against RGB stars in open clusters, and because of their larger dynamic range in parameter
space --- radius and luminosity, in particular --- compared to RC stars.

\subsection{The {\it Gaia} Data Release 2 sample}
The DR2 catalogue contains a host of useful astrometric, photometric,
and derived quantities for our purposes. As described in \cite{lindegren+2012a}, the global {\it Gaia} astrometric solution is an iterative process that proceeds in stages: first, the astrometric quantities for each star --- including the parallaxes, $\varpigaia$, that we use in this work --- are updated by minimizing the difference between the observed and predicted locations of the source images on the detector; next, the parameters describing the pointing of the {\it Gaia} satellite are updated; then the calibration solution is improved, which describes how the observed positions of the sources are systematically affected by instrumental effects like CCD irregularities, mechanical variations, and thermal fluctuations.\footnote{A final step allows for, e.g., General Relativity variations.}

Chromatic effects can affect the position of a source on the detector, meaning that there are generally color-dependent offsets in the observed position of a star on the detector that should be accounted for in the calibration part of the solution. Although not a part of the {\it Gaia} DR1 calibration solution, the global astrometric solution described in DR2 includes an additional term in the calibration step of the solution that depends on a proxy for color, $\nueff$. This quantity is the inverse of the effective wavenumber of a star, and depends on its spectral shape. $\nueff$ would normally be computed through an effective wavenumber-color relation using $G_{BP}$ and $G_{RP}$, for instance (Equation 2 in L18). However, given an initial calibration solution that describes how chromaticity affects the positions of stars on the detector, $\nueff$ can be estimated by adding it to the astrometric part of the solution (see \S3.1 of L18). For stars where a five-parameter astrometric solution is possible, this astrometric pseudo-color is reported in {\it Gaia} DR2, and has units of inverse micrometers. $\nueff$ tends to have more information about instrumental effects than would an effective wavenumber computed from photometry, as indicated by observations by \cite{arenou+2018} that parallax systematics correlate more strongly with $\nueff$ than with {\it Gaia} color, $G_{BP} - G_{RP}$. We therefore use $\nueff$, as an explanatory factor in our model to describe the
offset between asteroseismic and {\it Gaia} parallaxes. 

We also make use of {\it Gaia} DR2 photometry, including {\it Gaia} $G$, and the blue and red bandpass photometry, $G_{BP}$ and $G_{RP}$. The photometry is reduced based on the positions of the sources from the global astrometric solution, and internally calibrated according to \cite{riello+2018a}.

We only use stars in common with APOKASC-2 and DR2 by matching on 2MASS
ID, and from those, only keep those that meet criteria used
by \cite{andrae+2018}, namely:

\begin{itemize}
  \item \texttt{astrometric\_excess\_noise} = 0
  \item \texttt{visibility\_periods\_used} > 8
\end{itemize}
and with $\chi^2 \equiv$ \texttt{astrometric\_chi2\_al}, $n \equiv$
\texttt{astrometric\_n\_good\_obs\_al} - 5, $G_{BP} =
\texttt{phot\_bp\_mean\_mag}$, $G_{RP} = \texttt{phot\_rp\_mean\_mag}$,
\begin{itemize}
  \item $\chi \equiv
    \sqrt{\chi^2/n}$, $\chi < 1.2\mathrm{max}(1, \exp{-0.2(G -
      19.5)})$
  \item $1.0 + 0.015(G_{BP} - G_{RP})^2 <$
    \texttt{phot\_bp\_rp\_excess\_factor} $< 1.3 + 0.06(G_{BP} - G_{RP})^2$
\end{itemize}

These quality cuts ensure a good astrometric solution. We also
exclude a handful of stars whose parallaxes or radii that we derive below
disagree between asteroseismology and {\it Gaia} at the $5\sigma$ level. We do not
explicitly exclude negative parallaxes, and our analysis method described in
the next section does not require positive parallaxes. However, after
the above cuts, only positive parallaxes remain.

\section{Methods}
\label{sec:methods}

A star's radius, $R$,
is related to its parallax, $\varpi$, through its effective temperature,
$\teff$, and its bolometric flux, $F$, via
\begin{align}
  \label{eq:r_plx}
  \varpi(\teff, F, R) &= F^{1/2} \sigma_{\mathrm{SB}}^{-1/2} \teff^{-2}
  R^{-1} \nonumber \\
  &= f_0^{1/2}10^{-1/5(m + BC(b, \teff) - A_b)} \sigma_{\mathrm{SB}}^{-1/2} \teff^{-2} R^{-1},
\end{align}
where $\sigma_{\mathrm{SB}}$ is the Stefan-Boltzmann constant; $f_0 = 2.5460\times10^{-5} erg/s/cm^2$ is a zero-point factor to convert magnitude to flux and is computed assuming a solar irradiance from \cite{mamajek+2015a}, $f_0 = 1.361\times10^{6} erg/s/cm^2$, and an apparent solar bolometric magnitude of $m_{\mathrm{bol}} = -26.82$ (using the visual magnitude of the Sun, $V_{\odot} = -26.76$, and its visual bolometric correction, $BC_{V,\odot} = -0.06$; \citealt{torres+2010a}); $BC$ is
the bolometric correction, which depends on the photometric bandpass
used, $b$, and the temperature; and $A_b$
is the extinction in that band. One may use this equation to compute a
radius from the {\it Gaia} parallax, or a
parallax from an asteroseismic radius. Asteroseismic
radii themselves are derived from a radius scaling relation using the asteroseismic properties $\dnu$
and $\numax$, which represent the
typical frequency spacings between acoustic overtone modes in solar-like oscillators and
the frequency for which those oscillations are largest:
\begin{equation}
\frac{R}{\rsun} \approx \left(\frac{\numax}{\numaxsun}\right) \left(\frac{\dnu}{\dnusun}\right)^{-2}\ \left(\frac{\teff}{\teffsun}\right)^{1/2}.
\label{eq:scaling3}
\end{equation}

The published {\it Gaia} radii depend on an estimate of the flux of the star,
and not just the parallax. Because the published {\it Gaia} radii are
computed without taking into account extinction, the published {\it Gaia} radii are
systematically too small. To remove this known effect, we calculate
our own radii using {\it Gaia}
parallaxes, according to Equation~\ref{eq:r_plx} with visual photometry, which we will refer
to henceforth as {\it Gaia} radii.

One can see that Equation~\ref{eq:r_plx} suggests that for our
comparison between {\it Gaia} and asteroseismic results, we can either use
{\it Gaia} parallaxes with a flux and a temperature to yield a {\it
  Gaia} radius,
or alternately use the asteroseismic radius along with a flux and temperature to compute a parallax. In the following, we consider both approaches.

\subsection{Parallax comparison}
\label{sec:plx_comp}
It is simplest to identify a
zero-point offset in the {\it Gaia} parallaxes in parallax
space---i.e., by converting asteroseismic radii into asteroseismic
parallaxes. The following equations represent our assumptions that the
\textit{observed} {\it Gaia} parallaxes are offset from the true
parallax, $\varpi$, by a constant, global value, $c$, and are subject to
Gaussian measurement/modelling
noise (Equation~\ref{eq:plx1}); and the \textit{observed} asteroseismic radii are unbiased
measurements of the true parallax, subject to Gaussian noise (Equation~\ref{eq:plx2}):
\begin{align}
  \hatvarpigaia &\sim \mathcal{N}(\varpi - c, \sigma^2_{\varpigaia})   \label{eq:plx1}\\
  \hatvarpiast &\sim \mathcal{N}(\varpi, \sigma^2_{\varpiast}),   \label{eq:plx2}
\end{align}
The variance due to measurement/modelling noise for the observed
{\it Gaia} parallaxes is taken from the
\texttt{parallax\_error} field of the {\it Gaia} catalogue; the
variance for the observed asteroseismic parallax is computed by
applying standard propagation of error to Equation~\ref{eq:r_plx},
thereby treating the fractional variance in asteroseismic parallax as the
appropriate weighted sum in quadrature of the fractional variances of 
flux, temperature, and asteroseismic radius. In our analysis, we ignore objects that were $5\sigma$ outliers in parallax difference.

Equations~\ref{eq:plx1}~\&~\ref{eq:plx2} propose that the
difference between asteroseismic and {\it Gaia} parallax scales is due
to a constant zero-point error in the {\it Gaia} parallaxes, like the
one found by the {\it Gaia}
team. Astrophysically, this is a reasonable model, given that errors
in the three pillars underpinning the asteroseismic parallax scale --- the scaling
relation radius, temperature, and bolometric correction --- result in
{\it fractional} and not additive errors in the
asteroseismic parallax (Equation~\ref{eq:r_plx}). It is for that reason
that we have treated the
random uncertainties in asteroseismic parallax fractionally. By extension, this means that in the presence of systematic
errors in the asteroseismic parallax, the observed parallaxes
would be fractionally different from the true parallax. This is to be
contrasted with the {\it Gaia} parallax case: L18 expect systematic
errors in the {\it Gaia} parallax to be additive, not fractional,
due to the nature of the mathematical degeneracy that L18
proposes may produce the constant, global {\it Gaia} parallax
zero-point error they find. Instead, global problems in the radius or
temperature scale, for instance, would be parallax- and therefore distant-dependent
effects.  Indeed, in the presence of systematic errors, the asteroseismic parallax would
generally be wrong by a fractional factor, $f$, meaning that
Equations~\ref{eq:plx1}--\ref{eq:plx2} would read:
\begin{align}
  \hatvarpigaia' &\sim \mathcal{N}(\varpi - c, \sigma^2_{\varpigaia})   \label{eq:plx1_}\\
  \hatvarpiast' &\sim \mathcal{N}(f\varpi, f^2\sigma^2_{\varpiast}),   \label{eq:plx2_}
\end{align}
And the observed parallax difference, $\hatvarpiast' - \hatvarpigaia'$ would not be described by a
simple additive offset, $c$, but rather a parallax-dependent function,
$(f - 1)\varpiast + c$.

Equations~\ref{eq:plx1_}~\&~\ref{eq:plx2_} suggest a test of our
assumption that the offset is due to {\it Gaia} and not
asteroseismology errors: parallax-dependent offsets could indicate
systematic errors in the asteroseismic
parallaxes. In the analysis to
follow, we therefore investigated the parallax
offset as a function of parallax \& other observables, and as a
function of different populations, which might indicate more subtle,
population-dependent asteroseismic parallax errors. We find no strong evidence for a
problem with either the radius scale or the temperature scale, and we
place limits on the effects of bolometric correction systematics, as
well. We thus take $c$ to be an estimate in the {\it Kepler} field of the global {\it Gaia} parallax
error found in \cite{lindegren+2018} and \cite{arenou+2018}.

\begin{figure}
\includegraphics[width=0.5\textwidth]{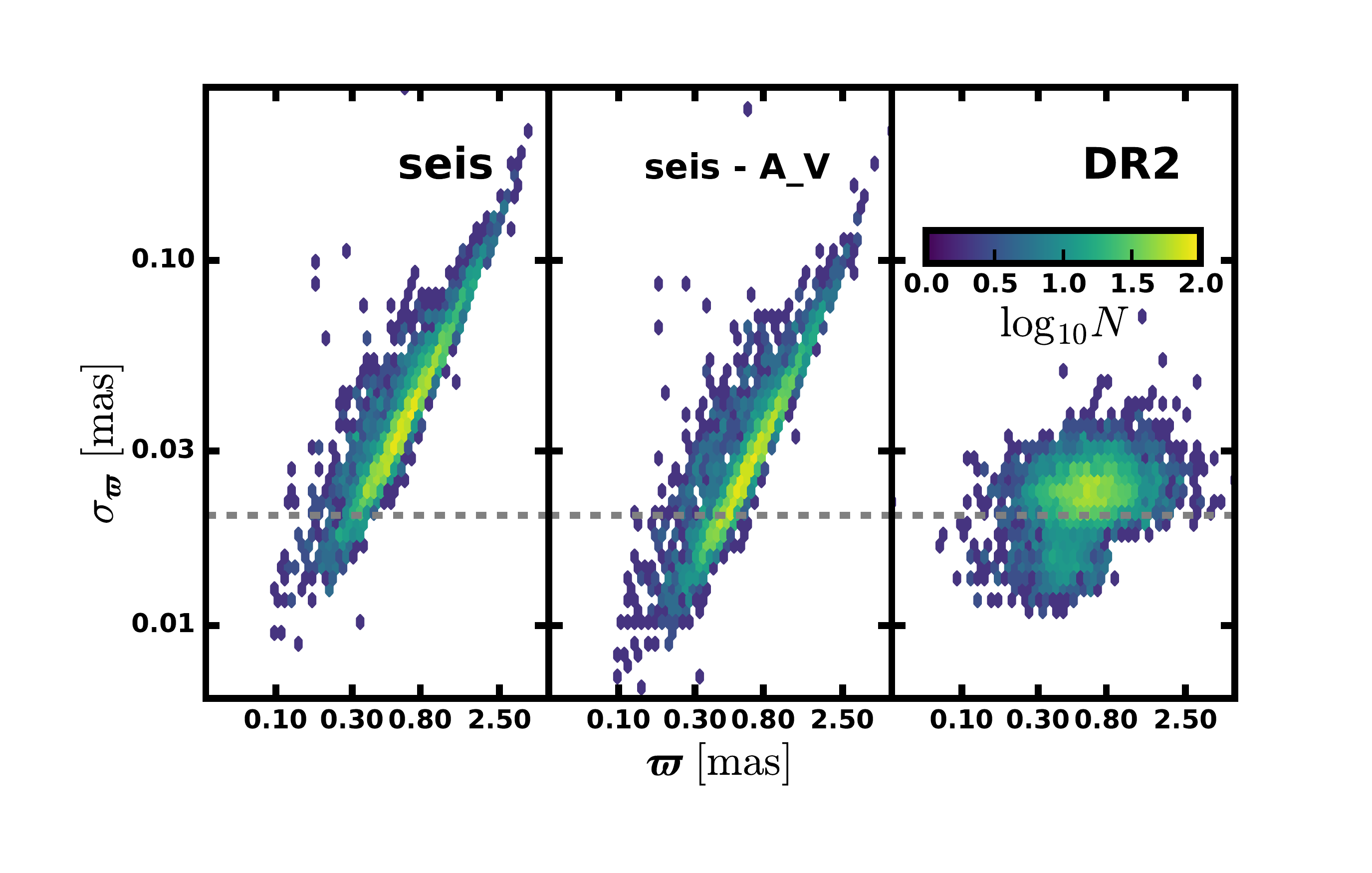} 
\caption{Parallax errors as a function of asteroseismic parallax with
  and without $A_V$ errors (``seis'', ``seis-A\_V'')
  and {\rm Gaia} parallax for our main sample. The color scale indicates the number of points per
  hexagonal cell. The horizontal dashed line indicates an error of
  $0.02\mas$, to guide the eye. Note the logarithmic scale on both axes.}
\label{fig:err}
\end{figure}

The dominant assumption in our analysis is that the asteroseismic
parallax/radius scale is the absolute one, and the {\it Gaia}
parallax/radius scale deviates therefrom. We examine in more detail this assumption that the asteroseismic parallaxes are \textit{accurate} in \S\ref{sec:discussion}. Apart from the matter of accuracy, our present analysis benefits from the exquisite precisions of our asteroseismic parallaxes. 
If the asteroseismic parallax precisions were worse than those from {\it Gaia}, the uncertainty on our inferred global offset $c$ would suffer. However, this is not the case. If taken at face value, Figure~\ref{fig:err} demonstrates that the asteroseismic parallaxes for our sample are at least as precise as that of the {\it Gaia} parallaxes, if not more. In fact, we have
reason to suspect that the {\it Gaia} parallax uncertainties are under-estimated. For stars with $G < 12$, the uncertainties appear well-behaved, thanks
to a post-processing inflation to the formal uncertainties the {\it Gaia}
team applied (L18). However, through
comparisons to literature distances, \cite{arenou+2018} has pointed out that parallax
uncertainties are significantly under-estimated, by as much as $40\%$ for $ 13 < G
< 15$ (the regime in which $15\%$ of our sample lives). For the purposes
of our analysis, we have assumed that the formal uncertainties on {\it Gaia}
parallaxes are accurate.  On the asteroseismic side, our asteroseismic parallaxes have individual precisions of about $5\%$, meaning that our entire sample of $\sim 3500$ stars naively could constrain $c$ to better than $0.1\%$ or $0.5\muas$. This may very well be an overestimate of our asteroseismic parallax uncertainties, as we have reason to suspect that the $A_V$
uncertainties are inflated, which increases the asteroseismic parallax uncertainties
(``seis'' versus ``seis - A\_V'' in Figure~\ref{fig:err}). Indeed,
we find no evidence for an increase of scatter in {\it Gaia} radius at
fixed asteroseismic radius for stars with larger formal $A_V$ uncertainties. We
investigate the impact of $A_V$ on our results in
\S~\ref{sec:syst_disc}, and find our result is robust across different
extinction regimes. For the purposes of this work, then, we
conservatively assume the Bayesian uncertainty estimates of
$A_V$ from \cite{rodrigues+2014}.

We find that the goodness-of-fit, as quantified by $\chi^2/dof$, of our inferred parallax offset
between asteroseismology and {\it Gaia} depends crucially on
allowing for spatially-correlated errors according to the estimate for
the spatial covariance matrix of the {\it Gaia} parallaxes in Equation~16 of
L18. We write the covariance of the {\it Gaia} and asteroseismic
parallax difference between two
stars $i$ and $j$ separated by an angular distance, $\theta_{ij}$, as
$C_{ij}(\theta_{ij}) = f(\theta_{ij}) +
\delta_{ij}\sigma^2_{\varpigaia,i} + \delta_{ij}\sigma^2_{\varpiast,i}$, where $f(\theta_{ij})$ describes the spatial
correlations in {\it Gaia} parallax error, and $\delta_{ij}$ is the Kronecker
delta function. L18, using their quasar reference
sample, find $f(\theta_{ij}) = a \mathrm{exp} \left ( -\theta_{ij}/b
\right )$, with $a = 135 \muas^2$ and $b = 14^{\circ}$. This covariance function was fitted across
the entire sky, and models well the
covariance at the largest scales. However, because we want to
characterize the zero-point offset in the {\it Kepler} field
specifically, we can ignore the covariance at the largest scales, and
only consider the covariance on scales smaller than the {\it Kepler}
field. We tried three approaches for quantifying the small-scale spatial
correlations: 1) Adopting the exponential form from L18 as is ($a =
135 \muas^2$ and $b = 14^{\circ}$); 2) Adopting the same exponential form from L18, but with $a = 1500
\muas^2$ and $b = 0.11^{\circ}$ (the angular scale is taken from fits
to TGAS parallax covariance from \cite{zinn+2017}, and $a$ is chosen to reproduce the smallest-scale
behavior in the observed covariance in L18); and 3) ignoring
spatial correlations altogether. We settle on the L18 covariance
function (1), as it yields the best goodness-of-fit, and consider the
average spread of the best-fitting $c$ among these
methods as a systematic error of $\pm 1\muas$ on $c$ due to the choice of spatial
covariance. This is a minimum estimate of the
systematic error on the offset, and we discuss additional
systematic errors on the offset due to the bolometric correction,
temperature scale, and radius scale in
\S\ref{sec:syst}.

Conveniently, the {\it Kepler} field is easily sub-divided into
small patches that correspond to the spacecraft ``modules'' that house
the CCDs on which a
star's image is recorded for a given quarter\footnote{The
spacecraft turns by $90^{\circ}$ each quarter, so the same star is found on one of
four modules over the 16-quarter {\it Kepler} mission}. We choose
therefore to consider the errors on the parallax for stars of a given
module to be independent of those of stars on every other module.
As we are ignoring correlations in parallax on the largest scales,
this is justified, and roughly amounts to truncating our covariance function
at angular scales larger than the module
size of $\sim 2.4^{\circ}$. Our results are not sensitive to the
details of the module-level truncation, which we discuss in \S\ref{sec:syst_disc}.

Ignoring correlations among the observables
($\teff$, $\dnu$, $\numax$, $A_V$, $g$, and $r$) yields a likelihood
function for $N$ stars on each module, $m$:
\begin{align}
  \label{eq:plx_model}
  \begin{split}
&\mathcal{L}_m(c | \hatvarpigaia, \hatteff,
\hatdnu, \hatnumax, \hat{A}_V,\\
&\hat{g},
\hat{r}, \hat{BC}) = \frac{1}{\sqrt{(2 \pi)^N |C|}} \\
&\exp{\left[-\frac{1}{2} (\vec{y} - \vec{x})^{\mathrm{T}}
  C^{-1} (\vec{y} - \vec{x})\right]},
\end{split}
\end{align}
where
\begin{align*}
\vec{y} &\equiv \hatvarpiast(\hatteff, \hatdnu,
\hatnumax, \hat{A}_V, \hat{g}, \hat{r}, \hat{BC}) \\
\vec{x} &\equiv \hatvarpigaia  +  c,
  \end{align*}
and where the flux has been computed using $g$ and $r$ in combination with $A_V$ and a V-band bolometric correction that depends on $\teff$
from \cite{flower+1996}, $BC(V, \teff)$, on which we assign a $3\%$
uncertainty. This is lower by one percent than the
formal uncertainty on the BC for a typical star in our sample, but the
precise uncertainty adopted on the BC does not significantly change our result. The conversion from $g$ and $r$ to a V-band magnitude is
taken from Lupton
2005\footnote{\url{http://www.sdss3.org/dr8/algorithms/sdssUBVRITransform.php}}. The
uncertainty introduced from the transformation is negligible compared to the
uncertainties on $A_V$, given the $1\%$ uncertainties on $g$ and $r$. In
addition to the random uncertainties of $A_V$, $g$, $r$, and
$\teff$, the asteroseismic parallax uncertainty,
$\sigma^2_{\varpiast}$, which enters on the diagonal of $C$ (see
discussion of parallax covariance, above), also incorporates the random uncertainties on $\dnu$ and $\numax$. In this way,
random uncertainties on both {\it Gaia} and asteroseismic
parallaxes are accounted for, as are spatially-correlated systematic errors on {\it Gaia}
parallaxes from the non-diagonal entries of $C$. We ignore other forms
of correlations among the observables that enter into the right-hand side of Equation~\ref{eq:r_plx}, for instance between $g$ and $r$; these are small corrections to the final uncertainty in the asteroseismic parallax --- for instance, accounting for the correlation between the $\teff^{-2}$ term in Equation~\ref{eq:r_plx} and the temperature dependence of asteroseismic radius, $R$ (Equation~\ref{eq:scaling3}), inflates the uncertainty on individual asteroseismic parallaxes by $\sim 10\%$, which negligibly impacts the inferred central values or uncertainties of our final result.

Because L18 points out color and magnitude dependences of the parallax
zero-point error among their comparison quasar sample, we build upon
this model by adding {\it Gaia} color and magnitude terms:
\begin{align}
  \label{eq:second_model}
  \begin{split}
&P_m(c,d,e | \hatvarpigaia, \hatteff,
    \hatdnu, \hatnumax, \hat{A}_V,\\
    &\hat{g}, \hat{r}, \hat{BC}, \hat{G}, \hatnueff) \propto \\
&\exp{\left[-\frac{1}{2} (c - \bar{c})^2/\sigma_c^2/M\right]} \\
&\frac{1}{\sqrt{(2 \pi)^N |C^{'}|}} \exp{\left[-\frac{1}{2} (\vec{y} -
  \vec{x})^{\mathrm{T}}  C^{'-1} (\vec{y} - \vec{x})\right]},
\end{split}
\end{align}
where
\begin{align*}
\vec{y} &\equiv   \hatvarpiast(\hatteff, \hatdnu,
      \hatnumax, \hat{A}_V, \hat{g}, \hat{r}, \hat{BC}) \\
\vec{x} &\equiv \hatvarpigaia  +  c + d(\hatnueff - 1.5) + e(\hat{G} - 12.2).
\end{align*}

This is our final model, which we assume for all our results unless
otherwise stated, and which describes the Bayesian posterior
probability of the parameters $\mu_m \equiv \{c, d, e\}$, for each
module, $m$, out of a total of $M=21$ modules. We use a prior on $c$ based
on our best-fitting value for the model with no color or magnitude
terms or spatial correlations in parallax, $\bar{c} \approx 55 \muas$, and a width approximately twice its
uncertainty, $\sigma_c \approx 1.5 \muas$, as reported in the next
section, though our results are insensitive to including the prior or
having an implicit, flat, improper prior for $c$. This prior is also used in the parallax space comparisons described in \S\ref{sec:syst}. Here, the covariance includes two additional terms along the
diagonal: $C^{'}_{ij} = C_{ij} + \delta_{ij}d^2\sigma^2_{\nueff,i} +
\delta_{ij}e^2\sigma^2_{G}$, with $\sigma_{\nueff}$ as the
\texttt{astrometric\_pseudo\_colour\_error} field for star $i$ from the DR2
catalogue, and $\sigma_G$ the uncertainty on $G$, which we
assign as $1\%$, which reflects the 10mmag-level systematics in {\it Gaia} photometry for $G > 3$ based on comparison to external catalogues \citep{evans+2018}. We assign the values $12.2$ and $1.5$, which are the medians of
$G$ and $\nueff$ for our sample, to center the magnitude- and
color-dependent terms. In this way, a star with the median $\nueff$ of
$1.5$ and the median G-band magnitude
of $12.2$ would have no magnitude or color correction. They are therefore not parameters in
this model.  We have used the astrometric source
color here as it should be more correlated with the astrometric
properties of the {\it Gaia} DR2 solution than $G_{BP} -
G_{RP}$. \cite{arenou+2018} indeed finds that the {\it Gaia} quasar parallax
zero-point is more sensitive to $\nueff$ than $G_{BP} - G_{RP}$ (see their Figure~18).

In what follows, $c$ will be referred to as a constant, global offset, to stress that it is a mean offset present in all {\it Gaia} parallaxes in the {\it Kepler} field.  The global term, $c$, is larger in magnitude for our sample than other, higher-order contributions to the {\it Gaia} parallax offset, which we model as color-, magnitude-, and spatially-dependent. Additional, color-, magnitude-, or spatially-dependent terms in the offset are taken into account in our model, but we attempt to make a distinction between the ``constant" or ``global" offset, $c$, that applies to all {\it Gaia} parallaxes in our sample, and the more general parallax offset appropriate for a particular star, given its color, magnitude, and position on the sky.

\begin{figure}[htb!]
  \centering
\includegraphics[width=0.5\textwidth]{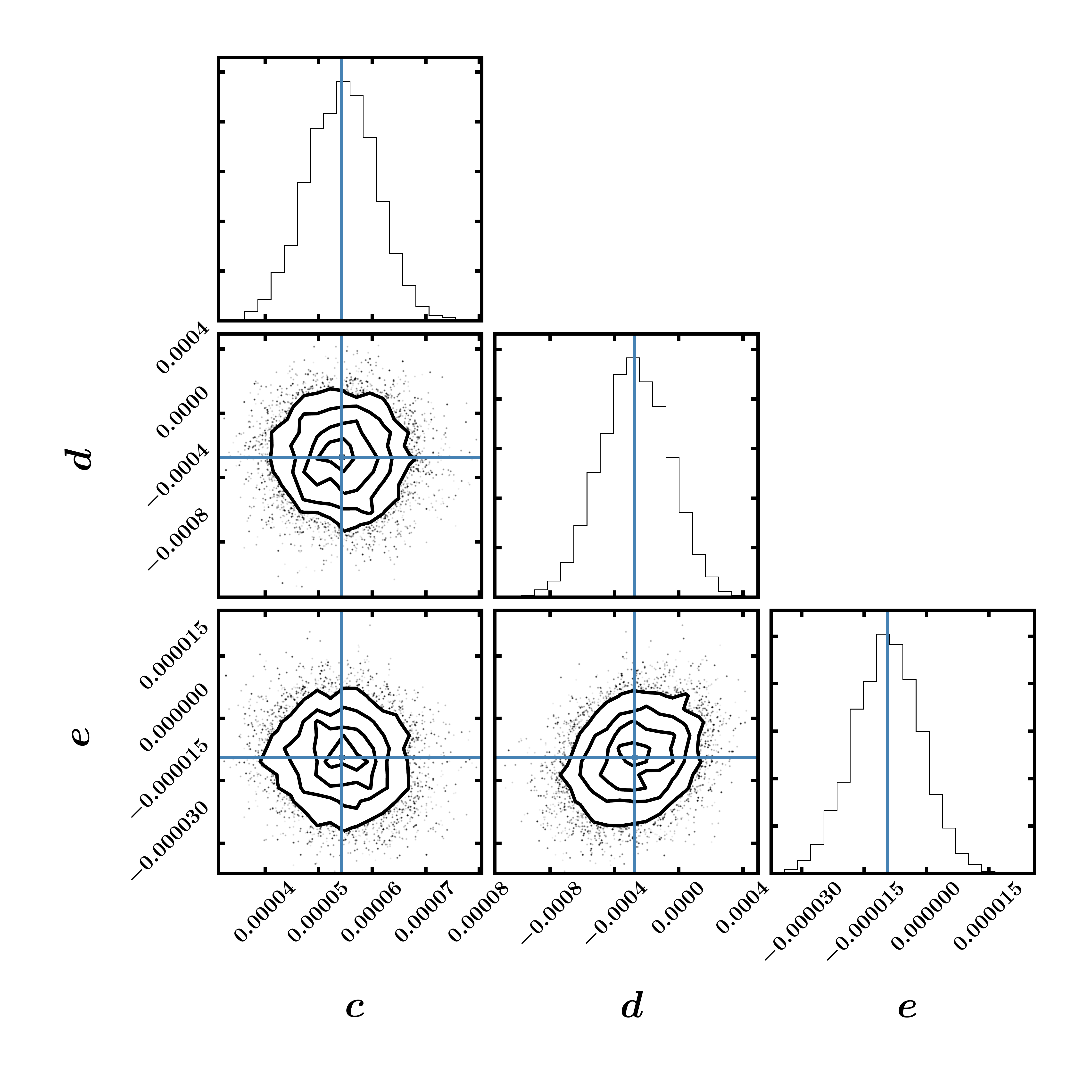}
\caption{The posterior distributions of the parameters for our parallax offset model (Equation~\protect\ref{eq:second_model}) for a single {\it Kepler} module. We combine the posteriors for each module to yield our final best-fitting parameters according to Equation~\ref{eq:combine}.}
\label{fig:corner}
\end{figure}

Because we can consider each
module independently, we estimate $c$ for each module, and combine
their values assuming they are described by a Gaussian around true
values, estimated to be $\hat{\mu}_M$, with covariance, $\hat{\Sigma}_M$, which we estimate as:
\begin{equation}
  P_M(c,d,e) \propto \Pi_{m=0}^{M} P_m(c,d,e) \propto \mathcal{N}(\hat{\mu}_M, \hat{\Sigma}_M),
  \label{eq:combine}
  \end{equation}
where
\begin{align}
  \hat{\Sigma}_M &= \left( \sum_{m=0}^{M}
  \hat{\Sigma}_m^{-1}\right)^{-1} \label{eq:mcmc} \\
  \hat{\mu}_M &= \hat{\Sigma}_M \left(\sum_{m=0}^{M}
  \hat{\Sigma}_m^{-1} \hat{\mu}_m\right), 
\end{align}
where each module's best-fitting parameters, $\hat{\mu}_m$, and
covariance matrices, $\hat{\Sigma}_m$, are estimated from MCMC chains using
\texttt{emcee} (\citealt{foreman-mackey+2013}; see
\citealt{neiswanger_wang&xing2013} for this and other more elaborate MCMC
  parallelization procedures). The posterior distributions for an example set of module-level parameters, $\hat{\mu}_m$, are shown in Figure~\ref{fig:corner}.
  
\subsection{Radius comparison}
\label{sec:inv_rad}
We also investigate the {\it Gaia} offset in inverse radius space. In
this case, the error budget is allocated differently because the {\it Gaia}
radius inherits uncertainty from combining $\varpigaia$ with the flux and
temperature to yield a radius.

We opt to work in terms of inverse radius, which is directly
proportional to parallax, thereby avoiding the bias and high variance
when converting the parallax to a
distance with $d = 1/\plx$. We refer the reader to \cite{bailer_jones2015} for thorough
discussions of the pitfalls in the naive (non-inverted) distance
approach. This inverted radius formalism also does not require the parallax to be positive, as a
negative parallax indicates a noisy estimate of a large radius (small
inverse radius). (In our sample, however, the negative parallaxes are
filtered out by the conditions listed in \S\ref{sec:data}). The formal
uncertainties on the asteroseismic radii $\sim 1.5 \%$, are such that inverting the
asteroseismic radius is well-tolerated.

We assume the observed {\it Gaia} radius is offset from the true
radius through its observed parallax, which is offset by $c$ from the true
parallax, $\plx$,
and that it is subject to measurement/modelling
noise (Equation~\ref{eq:inv_rad1}). As we note in the previous
section, the offset, $c$, is applied to the {\it Gaia} parallax
because we interpret it as a systematic error in the
{\it Gaia} parallax and not in the asteroseismic parallax. We also
assume the observed asteroseismic radius is distributed around the
true radius, $R^{-1}$, with known measurement/modelling noise (Equation~\ref{eq:inv_rad2}).
\begin{align}
  \hatinvrgaia &\sim \mathcal{N}(\hat{F}^{-1/2} \sigma_{\mathrm{SB}}^{1/2}
  \hatteff^{2}(\varpi - c), \sigma^2_{Gaia,R} + \sigma_{FT,i}^2) \label{eq:inv_rad1} \\
  \hatinvrscal &\sim \mathcal{N}(R^{-1}, \sigma^2_{\mathrm{seis},R}) \label{eq:inv_rad2},
\end{align}
where a hat denotes an observed quantity; the variance of the
inverse asteroseismic radius, $\sigma^2_{\mathrm{seis},R}$, is computed according to standard error
propagation applied to Equation~\ref{eq:scaling3}, thus
incorporating random uncertainty contributions from $\dnu$, $\numax$, and APOGEE $\teff$; the variance of the {\it Gaia} inverse
radius is computed according to standard propagation of error applied
to Equation~\ref{eq:r_plx}, and for clarity has been split into a
contribution due to fractional
uncertainties in the observed {\it Gaia} inverse radius due to flux,
temperature, and {\it Gaia} parallax, denoted
$\sigma^2_{Gaia,R}$, and a smaller contribution due to the offset, $c$,
denoted $\sigma_{FT,i}^2$, which is the
variance of the quantity $c\hat{F}^{-1/2}\sigma_{\mathrm{SB}}^{1/2}\hat{\teff}^{2}$. The {\it Gaia} radius is computed from the same photometry, BCs, extinctions, and temperatures that the asteroseismic parallax was in the previous section.

We formulate Equations~\ref{eq:inv_rad1}~\&~\ref{eq:inv_rad2} into a
likelihood for our $N$ stars:
\begin{align}
  \label{eq:rad_model}
    \begin{split}
&\mathcal{L}(c | \varpigaia, \hatteff,
\hatdnu, \hatnumax, \hat{A}_V, \hat{g},
\hat{r}, \hat{BC}) = \\
& \Pi_i (2\pi)^{-N/2} (\sigma_{FT,i}^2 + \sigma^2_{\mathrm{seis},R,i} +
\sigma^2_{Gaia,R,i})^{-1/2} \\
&\exp{\left[-\frac{1}{2} (x_i - y_i)^2/(\sigma_{FT,i}^2 + \sigma^2_{\mathrm{seis},R,i} +
\sigma^2_{Gaia,R,i})\right]},
\end{split}
\end{align}
where
\begin{align*}
  x_i &\equiv\hatinvrscali(\hatdnu, \hatnumax, \hatteff) \\
  y_i &\equiv\hatinvrgaiai(\hatvarpigaia, \hatteff, \hat{A}_V,
  \hat{g}, \hat{r}, \hat{BC}) + \\
  &c \hat{F}(\hatteff, \hat{A}_V,
  \hat{g}, \hat{r}, \hat{BC})^{-1/2} \hatteff^2.
\end{align*}

Again,
we ignore other correlations among observables, for instance, in
temperature. This model is used for validating our main model,
Equation~\ref{eq:second_model}. We do not include spatial correlations
in the parallax for this model, and
neither do we fit for color or magnitude terms in radius space.

In what follows, we report the uncertainty
in $c$, $d$, and $e$ from the diagonal of the parameter covariance matrix described
above (Equation~\ref{eq:mcmc}), $\hat{\Sigma}_M$, and the
best-fitting values from $\hat{\mu}_M$, except for
the radius space comparison offset, $c$, (Equation~\ref{eq:rad_model}), where we take the mean and standard deviation
from our MCMC chains.

\section{Results}
\label{sec:results}
No matter the method used, we find a consistent asteroseismology-{\it Gaia} parallax
offset for our {\it Kepler} RGB sample of $\approx 53
\muas$. Our main RGB sample yields an offset of $52.8 \pm 2.4 {\rm\ (rand.)}
\pm 8.6 {\rm\ (syst.)}
\muas$, with color and magnitude terms of $-150.7 \pm
22.7 \muas \mu m$ and $-4.21 \pm 0.77 \muas \mathrm{mag}^{-1}$. This result is consistent with that inferred from the
radius-based method ($c = 56.3 \pm
0.65 \muas$), when fitting $c$ without the color
and magnitude terms ($c = 52.9 \pm 0.35 \muas$), when no spatially-correlated parallax errors are used ($c
= 54.8 \pm 0.66 \muas$). We discuss our systematic error term below.

\begin{figure}[htb!]
  \centering
\includegraphics[width=0.5\textwidth]{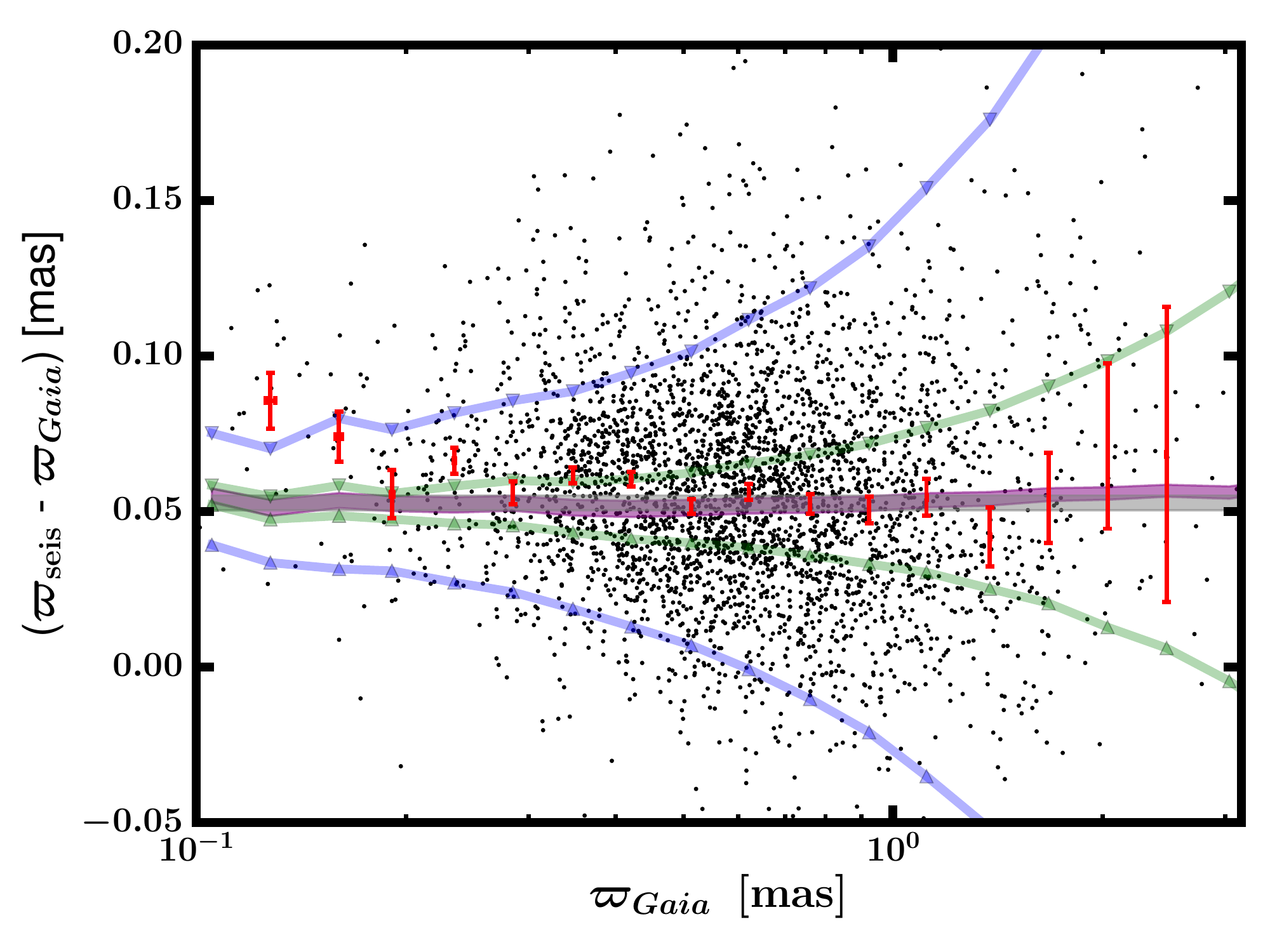}
\caption{Difference in {\rm Gaia} and asteroseismic parallax,
  as a function of {\rm Gaia} parallax. The observed data are shown in black, along with a binned median (red error bars). The $1\sigma$
  region for the best-fitting model using only a global offset of $c$ is indicated
  by the grey band, and
  one with color and
$G$ terms to describe the {\it Gaia} parallax offset by the purple band (with $\pm 1\sigma$
  in the global offset, and $\pm 0.5\sigma$ in the color and $G$
  terms).  Shown also
  are predicted effects from errors of $\pm 100K$ in the APOGEE temperature scale
  (blue) and $\pm 2\%$ in the radius
  scaling relation (Equation~\ref{eq:scaling3}; green). See text for
  details.}
\label{fig:plx}
\end{figure}

We visualize the offset in parallax as a function of both {\it Gaia}
parallax (Figure~\ref{fig:plx}) and asteroseismic parallax
(Figure~\ref{fig:rad_plx}b). In these figures, the grey band indicates a model for the parallax
offset being a constant equal to $c$, whereas the purple also takes
into account the color- and magnitude-dependent terms. Although it is
not evident in these plots, the
color- and magnitude-dependent terms are necessary to describe the
variability in the offset as a function of various observables, and
variations in the offset along these dimensions contribute to
the observed scatter away from the grey band in
Figures~\ref{fig:plx}~\&~\ref{fig:rad_plx}b. Indeed, it is only in looking at the offset as a function of our other
parameters in Figure~\ref{fig:tests} that we see the color and magnitude terms are required to
explain the data. This is particularly
evident of course in $\nueff$ and $G$ space
(Figures~\ref{fig:tests}e \&~\ref{fig:tests}f), but also notably
in $\dnu$ and $\numax$ space (Figures~\ref{fig:tests}b \&~\ref{fig:tests}c), where the color and magnitude terms
perform better than the global offset. The color term also shows up in the more familiar photometric color space, $G_{BP} - G_{RP}$, shown in Figure~\ref{fig:tests}h. The up-tick
of the offset for $ \varpigaia \lesssim 0.2\mas$ seen in
Figure~\ref{fig:plx} is likely due to
a bias in binned parallax values in the presence of large fractional
{\it Gaia} parallax error \citep[F. Arenou, personal communication;
][]{arenou+1999a}.

\section{Discussion}
\label{sec:discussion}
\subsection{Systematic errors in the zero-point offset}
\label{sec:syst}
Here, we present an estimate for the systematic error on our inferred {\it Gaia}
parallax offset, $c$, due to systematic errors in the asteroseismic
parallax scale we adopt. We want the systematic error to reflect how
accurate our reported parallax offset, $c$,
is in an absolute sense. As mentioned in \S\ref{sec:methods}, three
fundamental scales underpin the asteroseismic parallax: a bolometric
flux scale set by the bolometric correction; a temperature scale; and a
radius scale. Regarding the radius scale, the
APOKASC-2 asteroseismic radii have a $1\sigma$ systematic error of
$0.7\,\%$ due to
the uncertainty in the dynamical open cluster giant mass value that \cite{pinsonneault+2018}
adopts; this systematic error of $0.7\,\%$ in the radius scale
naturally accounts for systematic errors in both
$\numax$ and $\dnu$. Regarding temperatures, the IRFM scale, against which APOGEE temperatures are calibrated
\citep{ghb09}, agrees with the previously largest application of the IRFM
to giants by \cite{alonso+1999a} to within $\approx 43 K$, for the
metallicity range in which the majority of our sample lies ($-0.4 <
[Fe/H] < 0.4$). We
therefore adopt $43 K$ as a $2\sigma$ systematic error in the temperature scale.  Finally, as we note in \S\ref{sec:fidelity}, our bolometric correction choice
induces a systematic error of $5\muas$ on the inferred offset. A final
source of uncertainty is due to the {\it Gaia} spatial correlation
form, which we take as $1 \muas$ in \S\ref{sec:plx_comp}. Taken
together, these sources of systematic uncertainty on $c$ for the median
giant in our sample with $\plx = 600\muas$ and a temperature of $4700
K$ add fractionally in quadrature to produce an uncertainty in $c$
consistent with how parallax scales with radius and temperature
(Equation~\ref{eq:r_plx}):
\begin{equation*}
  \begin{split}
    \sigma_c &= 600 \muas \times \\
    & \left[(0.007)^2 + (43K/4700K)^2 + (5 \muas/600 \muas)^2 + (1
      \muas/600 \muas)^2\right]^{1/2} \\
&= 600 \muas \times 0.014 = 8.6\muas.
\end{split}
\end{equation*}

We therefore adopt a systematic error on $c$ of $8.6
\muas$. We argue in \S\ref{sec:fidelity} that temperature and
radius systematics are not visibly present in the data, and so this
systematic error estimate may be conservative.

More subtle errors in our inferred $c$ may arise due to population effects. For instance,
one may worry that our result could be biased by the distribution of
our sample's parallaxes and/or parallax uncertainties. We also are interested
in quantifying how sensitive our results are on the evolutionary state
of our sample. An obvious test is to analyze RC stars and compare to our
RGB results. Moreover, we would expect certain population effects to
map into a spatial dependence in our result. An age and/or metallicity
gradient in the distance above the Galactic disk could map out a spatial gradient in our
asteroseismic parallaxes, for instance. Extinction is also a strong
function of height above the disk, which could affect our inferred
fluxes in a spatially-dependent way. So although we take into account
spatially-correlated errors in {\it Gaia} parallax, we now place
limits on any spatial correlations in the asteroseismic parallaxes themselves. 

For these reasons, we performed several checks of the offset for different populations in
order to estimate any systematic effects biasing our inferred value of
$c$, including:
\begin{enumerate}
\item A high-parallax sub-sample with $\varpigaia > 1 \mas$
\item A high-precision sub-sample with $\sigma_{\varpigaia}/\varpigaia < 0.05$
\item Two high-extinction sub-samples (one with $\ell < 73^{\circ}$ and another with $b < 15^{\circ}$)
\item Two low-extinction sub-samples (one with $\ell >  73^{\circ}$ and another with $b > 15^{\circ}$)
\item A sub-sample consisting only of red clump stars (RC)
\item A metal-rich ($\mathrm{[Fe/H]} > 0.2$) and a metal-poor ($\mathrm{[Fe/H]} < -0.2$) sub-sample 
\end{enumerate}

In all of these cases, our parallax space model was used to
infer $c$ (Equation~\ref{eq:second_model}), and only RGB stars were
included (except for the RC sub-sample, which consisted exclusively of
RC stars). The results of the offsets and corresponding reduced $\chi^2$ are
tabulated in Table~\ref{tab:tests}. For comparison, our main sample is
included as ``RGB''. The agreement among all these methods is
excellent, and we discuss the implications for this agreement across
position on sky, extinction, and parallax in the next section.

\subsubsection{Population effects}
\label{sec:syst_disc}
From Table~\ref{tab:tests}, the only difference of note in the inferred
global offset from our fiducial RGB sample is in
the high $\varpigaia$ sub-sample, with a disagreement at the $1 \sigma$ level. The different inferred offset among the high $\varpigaia$
sub-sample can be explained by
the sub-sample's bright magnitude distribution, which requires a more
substantial magnitude correction term than the fainter RGB sample ($-13.25 \muas {\rm mag}^{-1}$
compared to $-4.21 \muas {\rm mag}^{-1}$).  Indeed, we do see evidence
for a non-linear magnitude-dependent parallax correction from
Figure~\ref{fig:tests}f, where the stars with $G < 10$ cannot be
described with a linear magnitude correction. The larger
magnitude-dependent parallax correction would then decrease the
difference in parallax scales for these bright, close stars compared
to the difference in parallax scales for the main RGB sample, and
result in a smaller inferred parallax offset, $c$. Note that this
implies a shortcoming in our linear magnitude-dependent correction,
and does not impugn the $c$ we infer from the broader RGB sample.

There is a statistically insignificant difference of $2.6\muas \pm 3.5
\muas$ between
the inferred offsets for the RGB and RC sub-samples, which indicates
that there are not large evolutionary state dependent offsets in the
derived asteroseismic parameters. The RC sub-sample does yield significantly different color and magnitude terms. We can understand the more negative $d$ in the RC sub-sample than in the RGB sample because the RC is bluer than the RGB, and so lies on the part of Figure~\ref{fig:tests}e where the color trend is non-linear and more severe ($\nueff \gtrsim 1.5$). The magnitude term being less negative than that from the RGB sample appears consistent with a scenario in which the red clump radii are too large by $\sim 1\%$, which would cause a biased trend in the parallax offset as a function of magnitude (green curves in Figure~\ref{fig:tests}f), and which would also result in a smaller $c$ among the RC population. This would be consistent with the systematic uncertainties in the RC radius scale being a factor of two greater than in the RGB radius scale \citep{pinsonneault+2018}, corresponding to a median systematic uncertainty of more than $2\%$. Given the robust calibration of radius for RGB stars, we
adopt the parallax zero-point offset estimated from the RGB
sample.

The other significant differences in the results of the sub-samples in Table~\ref{tab:tests} lie in a handful of color terms and magnitude terms differing from our RGB solution. Apart from the differing magnitude term for the high $\varpigaia$
sub-sample that we have already discussed, the color terms differ also for the $\ell < 73^{\circ}$ and  $\ell > 73^{\circ}$ cases. The reason for these differences appears to be that the reddest sources in our sample have a lower parallax offset for $\ell < 73^{\circ}$ and a larger one for $\ell > 73^{\circ}$. This means that the color trend is stronger in the $\ell > 73^{\circ}$ sub-sample because its red objects have a larger offset than the sub-sample with $\ell < 73^{\circ}$. The origin for this difference is unclear, but could be related to {\it Gaia} systematics in crowded regions, such as near the Galactic plane \citep{arenou+2018}. Additionally, the metal-rich sub-sample has a less negative color term, which is due to the sample being narrowly distributed in color space, along a relatively flat part of the color trend at $\nueff \approx 1.5$. Note that there are no metallicity-dependent effects on $c$, however, which indicates that there are small, if any, systematics in asteroseismic radii due to metallicity (Zinn et al., in prep.).

The too-low reduced $\chi^2$, $\chi^2/dof$, for some of the
sub-samples shown in Table~\ref{tab:tests} indicates
that the error budget for these sub-samples is inadequate, though we
do achieve an acceptable goodness-of-fit for
our main RGB sample. We note that
there is uncertainty in our result due to the unknown form of the
spatial covariance function for the {\it Kepler} field, which could
bias our reduced $\chi^2$ by changing the best-fitting value and/or
changing the effective number of degrees of freedom. We will
explore the latter effect in a future work.

Of particular interest is the consistency of the inferred offset in
high- and low-extinction regions. We can also see this visually as a flat trend
of the parallax difference as a function of extinction in
Figure~\ref{fig:tests}d. The APOGEE temperature scale we have adopted is spectroscopic, and
therefore insensitive to extinction, as are asteroseismic
frequencies. Fundamentally, then, the agreement across extinction regimes tells
us our combination of extinctions and bolometric correction yields
consistent distance estimates. In fields with larger reddening, however, a
$\ks$-band bolometric correction would likely be a better choice for
computing luminosities, given the $\ks$-band is insensitive to
extinction.

An interesting conclusion to draw from the different spatial
sub-samples we analyzed is the markedly low variation of the result
with spatial position. The four sub-samples
chosen in low- and high-extinction regions based on position with respect to
the Galactic plane agree to within $2.1\muas$, and this already small difference is
statistically insignificant. On the face of it, this
indicates not only that population effects on asteroseismic parallaxes
are quite small, but also indicates that the L18
prescription for spatial correlations in the {\it Gaia} parallaxes is
much smaller for our sample in the {\it Kepler} field. Instead of a
nominal parallax difference of $14\muas$ for separations of
$5^{\circ}$ according to Equation 16 of L18, they seem to be at most
at the $2 \muas$ level. This observation accords with our caution that
the L18 spatial covariance estimate is expected to be larger than one
expected for our sample, given the larger random uncertainties on {\it
  Gaia} parallaxes for the sample of quasars on which it is based,
which are much bluer, and five magnitudes fainter than our giant
sample.

To look further into the significance of lower than expected levels of spatial correlation,
we investigated to what extent there could be edge effects introduced by
truncating spatial correlations beyond the {\it Kepler} modules. By
not taking into account correlations between stars on neighboring modules, edge
effects may contribute to a biased $c$ or one with too-high inferred precision. We therefore analyzed four clusters of
three modules, which are located in each of the four corners of the {\it
  Kepler} field of view. The module clusters are separated from each other by
the width of a module, meaning these estimates of $c$ are unaffected by truncation
of spatial correlations at the module level. The mean $c$
we infer from these four clusters are $51.4\muas$, $57.0\muas$,
$56.6\muas$, and $52.8\muas$. These clusters deviate at most by
$4.2\muas$ from our final reported value of $52.8\muas$, which is well
within our systematic error budget of $8.6\muas$. This indicates that
our module-level truncation of the spatial correlations does not bias
our result or error budget, and confirms the markedly low spatial
variation on scales larger than $5^{\circ}$ we see among our extinction sub-samples.

\subsubsection{Fidelity of the APOGEE temperature scale and the radius
  scaling relation}
\label{sec:fidelity}
As we show in \S\ref{sec:methods}, a simple fractional modification
to our parallax model to describe a systematic asteroseismic radius error leads to a
parallax-dependent asteroseismology-{\it Gaia} parallax
difference. Here, we show that neither a radius error of this sort nor
a temperature error is consistent with the observed difference in
parallax scales. We therefore conclude that the observed parallax
difference is consistent with a global
systematic error in the {\it Gaia} parallaxes, with magnitude- and
color-dependent terms.

We show in Figures~\ref{fig:plx},~\ref{fig:rad_plx} and~\ref{fig:tests} what the
offset between asteroseismology and {\it Gaia} would be if the
APOGEE temperature scale differed by $+100K$ (blue curves with upward
triangles) and $-100K$ (blue curves with downward triangles), and if the asteroseismic radii were inflated by $2\%$ (green
curves with upward triangles) and deflated by $2\%$ (green curves with
downward triangles). The curves represent
the median deviation from the nominal, best-fitting model (purple
band) by perturbing the nominal temperatures or radii. In other words,
they indicate the behavior of the parallax difference in the presence
of significant systematic errors in the APOGEE temperatures or
radius scaling relation. These curves are indeed parallax-dependent, and commensurate
with the simple term in Equation~\ref{eq:plx2_}: 
temperature (blue curves) and scaling relation (green curves)
effects shown in Figure~\ref{fig:rad_plx}b, are fractional, and become more severe for larger
parallax. It is clear that none of these systematics can explain the
parallax offset that we infer, because Figures~\ref{fig:plx}~\&~\ref{fig:rad_plx}b show a
parallax offset that is essentially
flat as a function of parallax.

If Figures~\ref{fig:plx}~\&~\ref{fig:rad_plx}b demonstrate
that systematic errors in temperature or radius cannot cause the
observed parallax difference, more subtle, population-dependent
temperature or radius errors may still be present, which might be washed out when viewed as a function of only
parallax. In particular, we can use views of the same systematics curves shown
in Figure~\ref{fig:plx} in slices of other observables to evaluate to
what extent the color- and magnitude-dependent terms may be caused by systematic
errors in temperature or parallax. Figure~\ref{fig:tests} shows how these radius and temperature systematics play out as a function of the other
observables. These systematics curves do show very similar behavior to
the observed parallax difference as a function of the
observables. In each case, the systematics curves look consistent with
the observed parallax difference if they are shifted by a fixed, constant amount. None
of these curves, however, as we show in
Figures~\ref{fig:plx}~\&~\ref{fig:rad_plx}b, is consistent with the observed parallax offset
as a function of parallax, and so these translations are not
permissible. Nevertheless, for a relatively small number of stars on the
extremes of the parameter space, we could imagine these translations
are permissible, and would simply lead to larger scatter in the
parallax difference when viewed as a function of parallax alone. It is therefore interesting to compare the deviations in the
systematics curves at the extremes of some of these slices in parameter
space to the curvature in the observed parallax difference seen as a
function of color and magnitude. The
question in this scenario is whether temperature or radius systematics
can induce magnitude- or color-dependent trends in the observed parallax
difference.

Consider, as an example, the model for the effect of
an increase in the radius (green upward triangles), e.g., which can be
thought of as a situation where our radii are too large due to radius scaling
relation problems. Looking at the parallax difference in the
high-$\nueff$ regime in Figure~\ref{fig:tests}e, the offset between {\it Gaia} and
asteroseismic parallax actually decreases with increasing $\nueff$ because of a population effect, whereby stars with larger
$\nueff$ are closer, and thus a fractional increase in radius for
high $\nueff$ stars leads to a larger
absolute asteroseismic parallax shift --- in the case of too-high radii, their
parallaxes are shifted closer to the observed {\it Gaia}
parallaxes (Figure~\ref{fig:tests}a). This is the same sense of the
observed trend in parallax difference with $\nueff$, and could
therefore appear to contribute to a color term in the {\it Gaia}
parallax offset. However, an increase of the radius
scale is not consistent with the magnitude-dependent trend: whereas we
observe the nominal offset to
\textit{increase} for brighter stars, an increase of the radii results
in a \textit{decrease} in parallax offset for the brightest stars --- another population effect whereby
the brightest stars in APOKASC-2 are the closest (largest parallax)
--- meaning that a
fractional increase in radius for these stars leads to the largest
absolute parallax shift. We conclude that unexplained trends for $G < 10$
and $G_{BP}- G_{RP} < 1.2$ are more consistent with non-linear color- and
magnitude-dependent {\it Gaia} parallax systematics than temperature or radius systematics.

These same parallax-dependent trends in the systematics curves also appear
in Figure~\ref{fig:tests_k}, which is analogous to Figure~\ref{fig:tests}, but showing the parallax
difference when computing the asteroseismic parallax with a
$K_{\mathrm{s}}$-band bolometric correction. The BC is interpolated
from MIST \citep{dotter+2016a,choi+2016a} BC tables in metallicity,
gravity, and temperature. The tables are constructed using the stellar evolution code MESA \citep{paxton+2011a,paxton+2013a,paxton+2015a}, combined
with the C3K grid of 1D atmosphere models (Conroy
et al., in prep; based on ATLAS12/SYNTHE;
\citealt{kurucz+1970a,kurucz+1993a}). We use the asteroseismic gravity,
temperature, and metallicity from the APOKASC-2 catalogue for this BC. We see that the effects
of a systematic temperature error are significantly reduced compared
to the visual bolometric extinction, because for giants, the
$K_{\mathrm{s}}$-band is on the linear part of the blackbody curve: a
large change in temperature will not cause an exponential change in
the infrared flux. This bolometric correction approach is also less
sensitive to the estimate of the extinction due to decreased dust
scattering in the infrared compared to visual wavelengths. The offset
we find using this approach yields a value of $42.4 \pm
3.5\muas$. Unlike the case when we use a $V$-band BC, the $\ks$-band
BC appears to produce a parallax-dependent parallax difference that
could indicate the presence of a small systematic error in the $\ks$ BC. Were we to
model this error with a fractional term of the sort we propose
in our toy model for systematic radius or temperature errors (Equations~\ref{eq:plx1_}~\&~\ref{eq:plx2_}), we would
recover a parallax offset more consistent with the one found using the
$V$-band BC. This difference between $V$-band and $\ks$-band BC
approaches is therefore a conservative estimate of the error in the
parallax offset due to BC: allowing for a fractional term in the
parallax difference model would enable a more precise estimate of the
offset. We adopt the
difference in offset between the $K_{\mathrm{s}}$ BC and the $V$ BC,
$10 \mu as$, as a $2\sigma$ systematic error. We have included
this addition to the systematic error budget, along with systematic
errors due to the temperature scale and the scaling relation
radius scale in \S\ref{sec:syst}. That the offset inferred using
K-band photometry is very
similar to the one we infer with V-band photometry, even though the
V-band asteroseismic parallaxes are more sensitive to temperature
systematics (compare, e.g., the temperature systematics curves in
Figure~\ref{fig:tests_k} versus Figure~\ref{fig:tests}), further
supports our conclusion that temperature systematics are small. Furthermore, because infrared photometry is insensitive to extinction, the $10 \mu as$ $2\sigma$ systematic uncertainty we infer from this test also accounts for systematics or correlations in our adopted extinctions. 

Looking at the trends in all of
these dimensions in this way, there does not seem
to be evidence for significant problems in the temperature or APOKASC-2
radius scale that would cause either the global zero-point offset or the magnitude- or color-dependent {\it
  Gaia} terms we
infer. It is nonetheless possible that much smaller temperature offsets ($\sim
10K$) could exist than we have considered, and still be consistent
with the flatness in the parallax offset seen in Figures~\ref{fig:plx}~\&~\ref{fig:rad_plx}b. Regarding small radius systematics, based on work in preparation (Zinn et al., in prep.), there seems to be a small systematic error in the asteroseismic radius scale such that asteroseismic radii are slightly larger than {\it Gaia} radii along some parts of the giant branch, and which is at a level consistent with the systematic error of $0.7\%$ discussed in \S\ref{sec:syst}. This systematic is in the correct sense to explain why the inferred $c$ using a large parallax sample is marginally smaller than $52.8\muas$ (Table~\ref{tab:tests}). It is also
possible that radius and temperature scale systematics
could both be present, operating in different senses to shift the
inferred zero-point by a small amount while cancelling
the parallax-dependent offset behavior seen in the systematics curves. In any event, the systematic error in our inferred
offset accounts for such possibilities.
\subsection{Possible evidence for evolutionary-state--dependent radius scaling relation errors}
As a final word on the matter, we note that the above statements have assumed that any changes to the observed
parallax offset by radius scaling relation errors are present at all
radii (a constant fractional error). However, breakdowns in the
scaling relations are expected to occur for the most evolved giants \citep{mosser+2013,stello+2014}. The largest sample of giants with
asteroseismic and eclipsing binary masses and radii (10) indicate
evidence for scaling relations yielding inflated masses and radii
\citep{gaulme+2016}, though other results with smaller sample sizes
have shown consistency in mass and radius scaling relations \citep{frandsen+2013,brogaard+2016}. Moreover, interferometric radii for four giants in 
\cite{huber+2012} do not show evidence for systematics in the scaling
relations. It is therefore of great interest to test the fidelity of
scaling relations for evolved giants.

\begin{figure*}[htb!]
\centering
\includegraphics[width=0.48\textwidth]{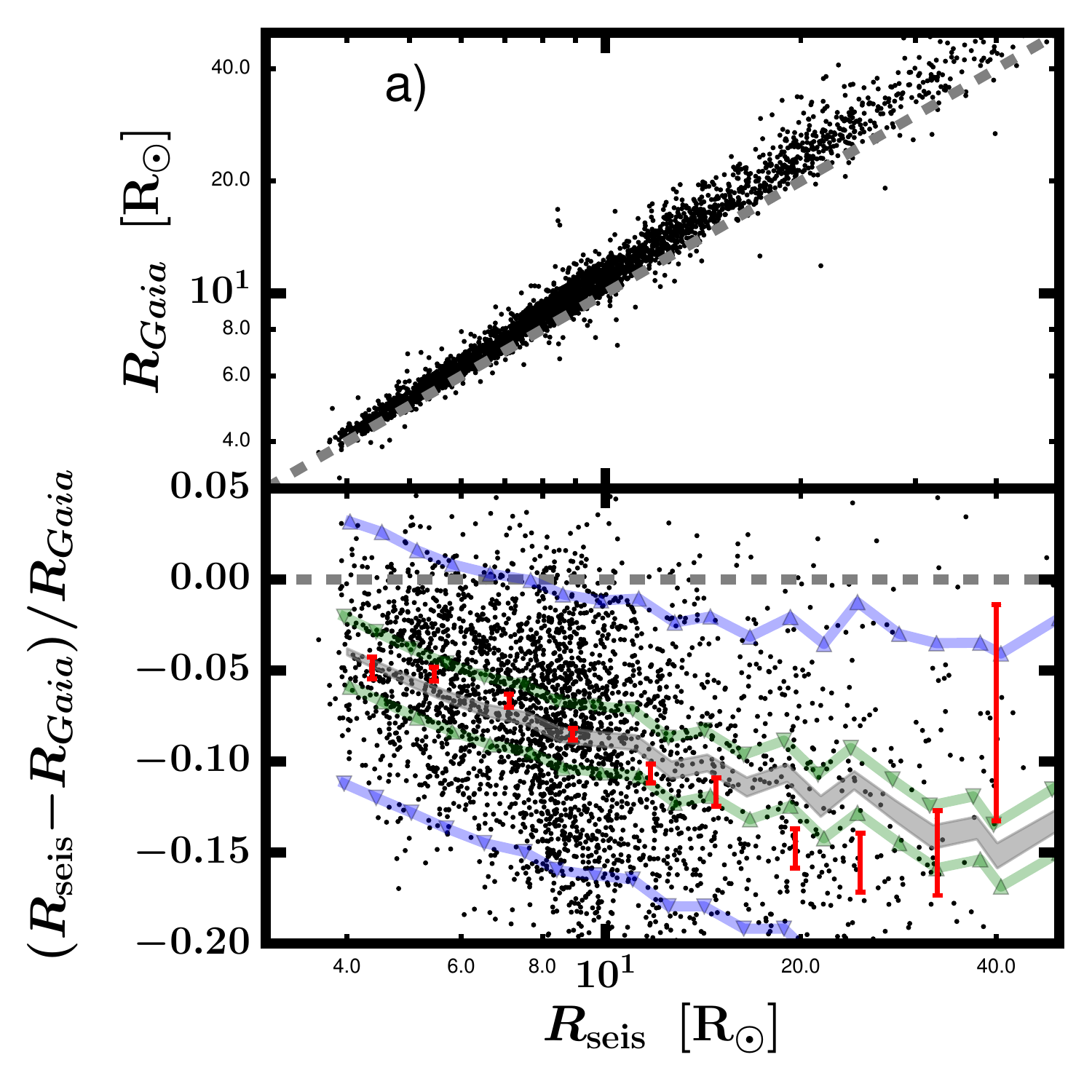}
\includegraphics[width=0.48\textwidth]{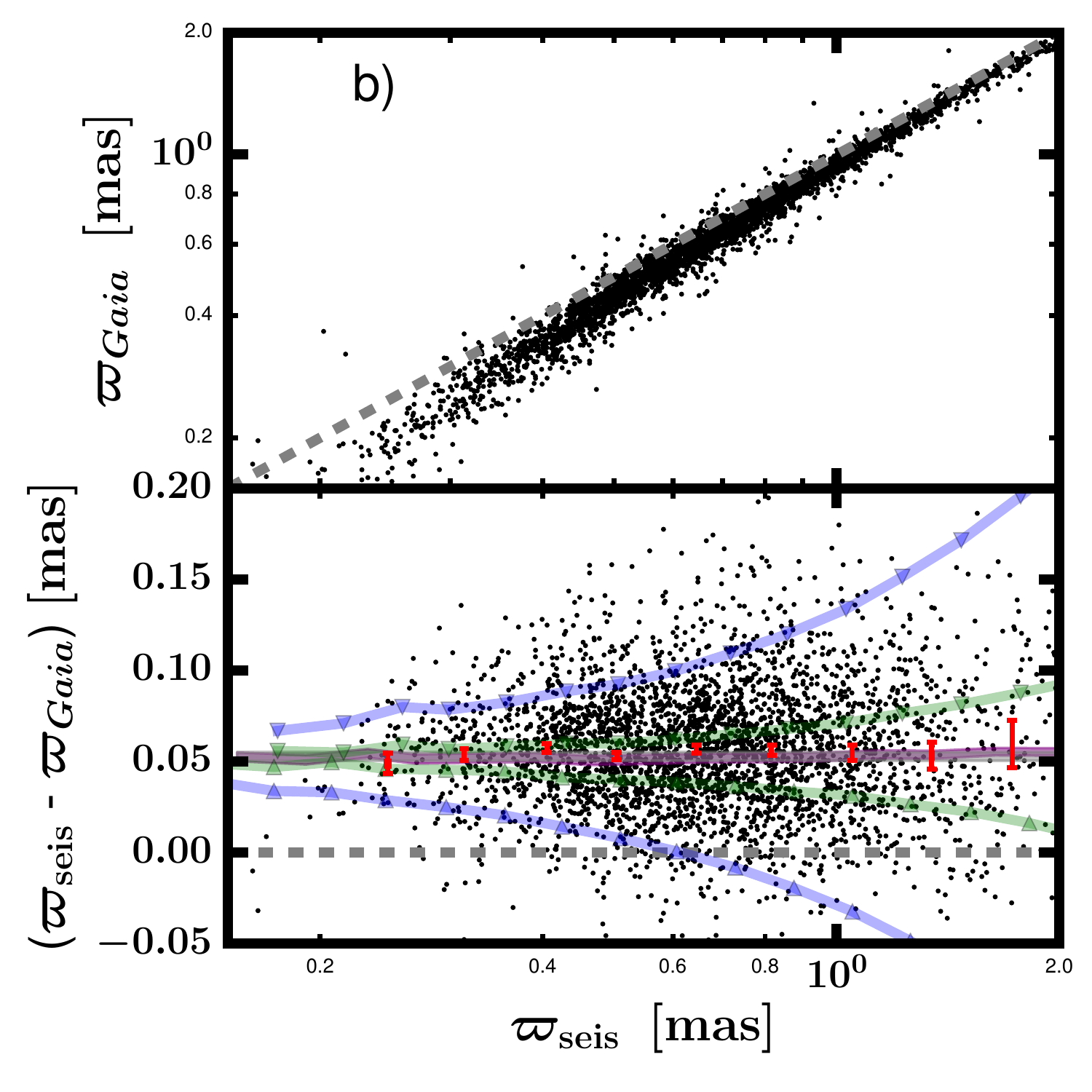}
\caption{Fractional difference in {\rm Gaia} and asteroseismic radii,
as a function of asteroseismic radii (a) and difference in {\rm Gaia} and asteroseismic parallax,
  as a function of asteroseismic parallax (b). In the radius
panel (a), the grey band
indicates the best-fitting
1$\sigma$ model, which only allows for an offset in the {\rm Gaia}
parallax (Equation~\ref{eq:rad_model}). In the parallax panel, the purple band indicates a model that allows for color and
$G$ terms in the {\rm Gaia} parallax offset (Equation~\ref{eq:second_model}),
with $\pm 1\sigma$ in the global offset, and $\pm 0.5\sigma$ in the
color and $G$ terms. The observed data (black) and
  binned median (red error bars) are
  well-described by a global offset of $c$ (grey band). Shown also
  are models describing how the binned medians of the data (red error
  bars) would appear in the presence of systematic errors of $\pm 100K$ in the APOGEE temperature scale
  (blue) and systematic errors of $\pm 2\%$ in the radius
  scaling relation (Equation~\ref{eq:scaling3}; green). See text for details.}
\label{fig:rad_plx}
\end{figure*}

\begin{figure*}[htb!]
\includegraphics[width=\textwidth]{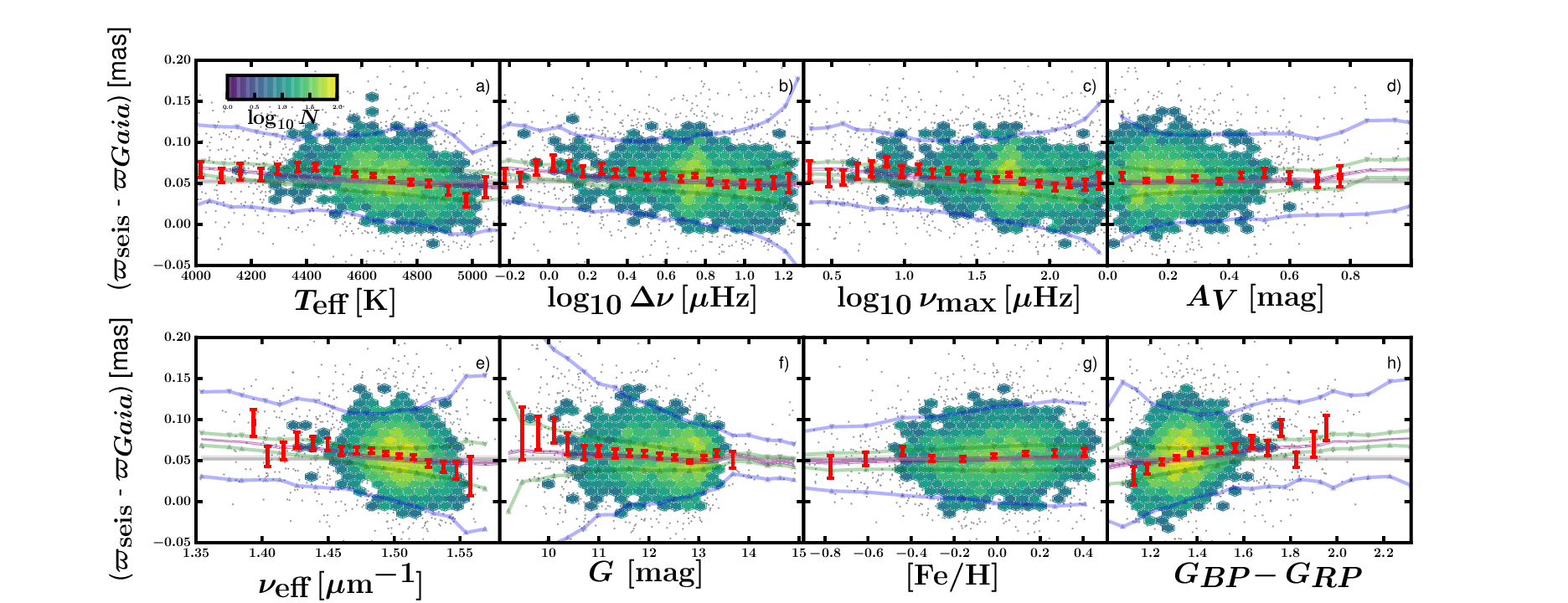}
\caption{Difference in {\rm Gaia} and asteroseismic parallax,
  as a function of $\teff$ (a), $\dnu$ (b), $\numax$ (c), $A_V$ (d),
  $\nueff$ (e), $G$ (f), $\mathrm{[Fe/H]}$ (g), and $G_{BP} - G_{RP}$ (h). In general, the observed data (black) and
  binned median (red error bars) are
  well-described by a global offset of $c$ (grey band), but better described by a model that allows for color and
$G$ terms in the {\rm Gaia} parallax offset (purple band; with $\pm 1\sigma$
  in the global offset, and $\pm 0.5\sigma$ in the color and $G$
  terms).  Shown also
  are predicted trends due to errors of $\pm 100K$ in the APOGEE temperature scale
  (blue) and $\pm 2\%$ in the radius
  scaling relation (Equation~\ref{eq:scaling3}; green). See text for
  details. Uncertainties on G-band magnitude
  have been set to $1\%$. The density of points per bin is denoted by the
  color bar in panel a.}
\label{fig:tests}
\end{figure*}

\begin{figure*}[htb!]
\includegraphics[width=\textwidth]{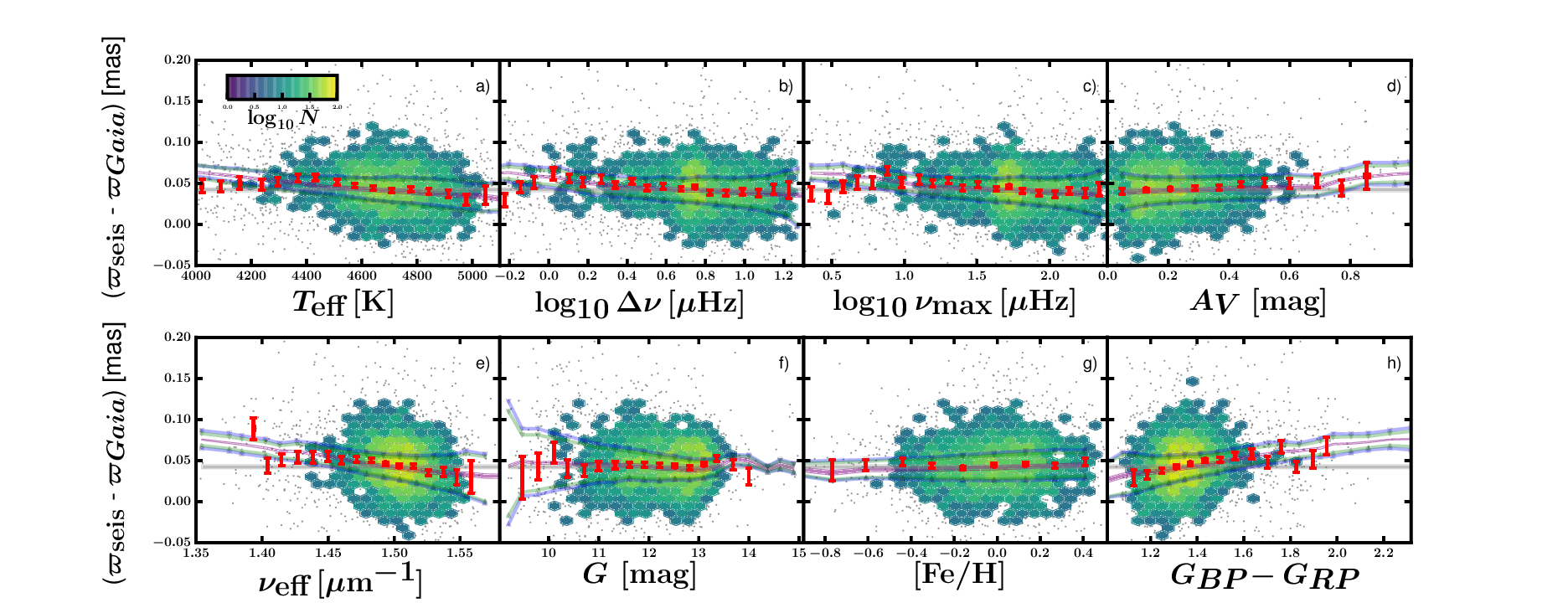}
\caption{The same as Figure~\ref{fig:tests}, except using a
  $K_{\mathrm{s}}$ bolometric correction when computing the
  asteroseismic parallax.}
\label{fig:tests_k}
\end{figure*}

In our data, a breakdown in the
radius scaling relation for evolved giants would manifest as a decrease in the parallax offset for low-$\dnu$ and
$\numax$ stars. We do see hints of this in Figures~\ref{fig:tests}b~\&~c,
where there is a down-tick in the parallax difference for $\dnu
\lesssim 1\muhz$ and $\numax \lesssim 10\muhz$. Given the expectation
that larger giants would have a stronger breakdown in the radius scaling relation, it might
also explain the flattening of the slope in fractional radius difference
above $\rscal \sim 20\rsun$, compared to our model (Figure~\ref{fig:rad_plx}a). However, because larger stars are cooler, this
effect may be degenerate with a non-linear {\it Gaia} parallax error,
where indeed we see that our linear model in color space cannot
simultaneously describe the data for cool stars with $\nueff \lesssim
1.45 \mu m^{-1}$ and the hotter stars with $\nueff \gtrsim 1.45 \mu
m^{-1}$ in
Figure~\ref{fig:tests}e; the linear color term also cannot precisely
fit the whole trend in temperature space in Figure~\ref{fig:tests}a. We cannot discriminate at this point between
a scaling relation problem or a color-dependent {\it Gaia} parallax offset term
that our linear model cannot describe. We emphasize, however, that any
hints at a scaling relation
breakdown only concerns interpreting small residuals at the extremes
(in color, radius, distance, etc.) of our APOKASC-2 parameter
space. We stress that the global {\it Gaia} parallax offset we report
is not consistent with being caused by temperature or scaling relation
errors, and neither are the main trends in color and magnitude in the regimes
where the bulk of our data lie.

\begin{table*}
  \begin{tabular*}{\textwidth}{cccccc}
    Sample & $c$ [$\muas$] & $d$ [$\muas \mu m$] & $e$ [$\muas
      \mathrm{mag}^{-1}$] & $\chi^2/dof$  & $N$ \\  \hline
         $b < 15^{\circ}$ & $53.7 \pm  2.8$ & $-125.5 \pm  26.4$ & $-4.10
    \pm
    0.89$ & $0.946^{*}$ & $ 2532$ \\ \hline
     $\ell < 73^{\circ}$ & $51.6 \pm  4.4$ & $-34.9 \pm  39.2$ &
    $-3.56
    \pm 1.41$ & $1.092^{*}$ & $  891$ \\ \hline
     $b > 15^{\circ}$ & $53.3 \pm  3.2$ & $-191.8 \pm  42.2$ & $-4.41
    \pm    1.48$ & $1.116^{**}$ & $  942$ \\ \hline
     $\ell > 73^{\circ}$ & $53.1 \pm  2.6$ & $-203.8 \pm  27.3$ &
    $-4.57  \pm 0.90$ & $0.975^{}$ & $ 2584$ \\ \hline
    $\varpi_{Gaia} > 1\mas$ & $46.0 \pm  5.7$ & $-255.7 \pm 146.3$ &
    $-13.25 \pm 3.82$ & $0.583^{*****}$ & $  555$ \\ \hline
    $\sigma_{\varpi_{Gaia}}/\varpi_{Gaia} < 0.05$ & $53.1 \pm  2.3$ &
    $-215.3 \pm  39.7$ & $-5.92 \pm 1.00$ & $0.781^{*****}$ & $ 2640$
    \\ \hline
    RC & $50.2 \pm  2.5$ & $-315.2 \pm  49.3$ & $0.79 \pm 1.07$ &
    $0.666^{*****}$ & $ 2587$ \\ \hline
[Fe/H] > 0.2 & $55.3 \pm  1.4$ & $-60.2 \pm  57.0$ & $-3.20 \pm 2.17$ & $0.688^{*****}$ & $  587$ \\ \hline
 [Fe/H] < -0.2 & $54.7 \pm  1.4$ & $-160.0 \pm  50.5$ & $0.42 \pm 1.73$ & $0.977^{}$ & $  828$  \\ \hline
    RGB & $52.8 \pm  2.4$ & $-150.7 \pm 22.7 $ &$-4.21 \pm 0.77 $&  $    
    1.007^{}$  & $ 3475$ \\\hline
\end{tabular*}
\caption{Terms in the {\rm Gaia} parallax zero-point offset and their uncertainties inferred from various APOKASC-2
  populations when fitting with
  Equation~\ref{eq:second_model}, and with a prior centered around $c = 55\muas$, which does not significantly affect the inferred parameters. See text for details. Asterisks denote the level of
  discrepancy with the expected $\chi^2$ given the degrees of freedom,
  $N - 3 - 1$, with one asterisk for each $\sigma$ in the significance
  of the discrepancy, capped at $5\sigma$. Our preferred value is from
  our main ``RGB'' sample. We estimate a systematic error in $c$ of
  $\pm 8.6\muas$. See text for details.}
\label{tab:tests}
\end{table*}

\section{Conclusions}
With a sample of nearly $3500$ first-ascent giants in the APOKASC-2
catalogue, we infer a systematic zero-point in the {\it Gaia} parallaxes of
$\varpiast - \varpigaia = 52.8 \pm 2.4 {\rm\ (rand.)} \pm 8.6 {\rm\ (syst.)} - (150.7 \pm 22.7)(\nueff - 1.5) - (4.21 \pm 0.77)(G - 12.2)  \muas$, in the sense that {\it Gaia} parallaxes are too
small. All indications are that the
zero-point offset is position-, magnitude-, and color-dependent (L18), so
we do not advise to use our model out-of-the-box. Our work does,
however, serve as a useful reference for other studies needing to
account for the zero-point offset in their work.

We have confidence in our result because of agreement with
the global zero-point parallax error of  $29 \muas$ that L18 finds for
a sample of nearly 600,000 quasars from AllWISE \citep{secrest+2015},
in the same sense
that we find. Though our global offset is larger in an absolute sense than
that found by L18, it is consistent with the range of zero-point offsets between $\approx 10 -
100 \muas$ (also in the sense that {\it Gaia} parallaxes are too
small) noted by the {\it Gaia} team \citep{arenou+2018}. This quoted range was determined from a census that appealed to more than
200,000 stars from 29 sources, ranging from {\it Hipparcos}
\citep{vanleeuwen2007} to the spectrographic survey LAMOST
\citep{luo+2015}. Several independent studies have also corroborated our findings \citep{riess+2018,stassun+2018a,groenewegen+2018a,muraveva+2018a,graczyk+2019,leung+2019a,schonrich+2019a,xu+2019a,khan+2019,hall+2019}, and our global offset is formally statistically consistent with nine of these, while one finds a global offset lower by $2\sigma$ \citep{graczyk+2019}, and which, like the L18 results, is a zero-point that has been averaged over the whole sky. Regarding this point, we note that Figure 7 in L18, which shows the global
error in the AllWISE quasar sample as a function of ecliptic
latitude, suggests that the {\it Kepler} field, at $\sim
64^{\circ}$, should exhibit a higher error ($\sim 50\muas$) than the
rest of the sky. We
also appear to find the same sign in the magnitude- and
color-dependent offset terms suggested by Figure 7 of L18.

We leave the reader
with the following conclusions:
\begin{enumerate}
\item For studies using
{\it Gaia} parallaxes of populations of red, bright stars in the {\it Kepler} field, we
think our estimate of $\varpiast - \varpigaia = 52.8 \pm 2.4 {\rm\ (rand.)}
\pm 8.6 {\rm\ (syst.)} - (150.7 \pm
22.7)(\nueff - 1.5) -
(4.21 \pm 0.77)(G - 12.2)  \muas$ should be valid, given the various tests provided
in \S\ref{sec:discussion}. In our sample, which has a range of $\sim 0.2$
in $\nueff$ and spans $\sim 4$ magnitudes, the color and magnitude
terms are appreciable. We have assigned a systematic error of $\pm
8.6\muas$ on the global offset due to our bolometric
  correction, choice of spatial correlation form for {\it Gaia} parallaxes, and systematics in
  our temperature \& radius scales.
  \item The parallax offsets we infer are not consistent with being
    due to significant systematic errors in the temperature or radius scale used
    to compute the asteroseismic parallax, and are in agreement with
    both the global and the magnitude- \& color-dependent parallax
    errors reported by the {\it Gaia} team \citep{lindegren+2018,arenou+2018}. 
  \item There are only insignificant
differences in the {\it Gaia} zero-point offset due to extinction in
the {\it Kepler} field, and due to population effects (e.g., red giant
branch versus red clump).
\item Our spatial covariance model of the DR2
parallaxes in the {\it Kepler} field likely needs
revision, which we will quantify in future work. At this point, there
are uncertainties on the parallax zero-point offset due to not knowing the precise nature of the spatial
correlations, which are at least $\pm 1\muas$.
\item Small trends in the data that our {\it Gaia} parallax model does not explain are suggestive of the need
  for a non-linear treatment of the magnitude- and color-dependence of
  the {\it Gaia} parallax offset. However, these trends appear
  preferentially for $G < 10$ and $G_{BP} - G_{RP} < 1.2$, where
  relatively few stars exist in our sample, and as such, we leave such an advanced treatment
  to other work.
\end{enumerate}

In this work, we did not attempt to map out the fidelity of the radius
scaling relation, though this is in principle possible, given the
difference in trend that a parallax error and a radius error induce
on the data. We will investigate this and estimate the spatial dependence of {\it Gaia} parallax errors in
 a forthcoming paper on tests of scaling relations as a function
 of evolutionary state.
 \newpage
 
\acknowledgments
M.~H.~P. and J.~Z. acknowledge support from NASA grants 80NSSC18K0391 and
NNX17AJ40G. D.~H. acknowledges support by the National Science Foundation
(AST-1717000) and the National Aeronautics and Space Administration
under Grants NNX14AB92G and NNX16AH45G issued through the Kepler
Participating Scientist Program and the K2 Guest Observer Program.
D.~S. is the recipient of an Australian Research Council Future Fellowship (project number
FT1400147). Parts of this research were conducted by the Australian
Research Council Centre of Excellence for All Sky Astrophysics in 3
Dimensions (ASTRO 3D), through project number CE170100013.

This project was developed in part at the 2019 Santa Barbara Gaia Sprint, hosted by the Kavli Institute for Theoretical Physics at the University of California, Santa Barbara. This research was supported in part at KITP by the Heising-Simons Foundation and the National Science Foundation under Grant No. NSF PHY-1748958.

This work has made use of data from the European Space Agency (ESA)
mission
{\it Gaia} (\url{https://www.cosmos.esa.int/gaia}), processed by the
{\it Gaia}
Data Processing and Analysis Consortium (DPAC,
\url{https://www.cosmos.esa.int/web/gaia/dpac/consortium}). Funding
for the DPAC
has been provided by national institutions, in particular the
institutions
participating in the {\it Gaia} Multilateral Agreement.

Funding for the Sloan Digital Sky Survey IV has been
provided by the Alfred P. Sloan Foundation, the U.S.
Department of Energy Office of Science, and the Participating Institutions. SDSS acknowledges support and
resources from the Center for High-Performance Computing at the University of Utah. The SDSS web site
is www.sdss.org. SDSS is managed by the Astrophysical Research Consortium for the Participating Institutions of the SDSS Collaboration including the Brazilian
Participation Group, the Carnegie Institution for Science, Carnegie
Mellon University, the Chilean Participation Group, the French
Participation Group, Harvard Smithsonian Center for Astrophysics, Instituto de Astrofs{\'i}ca de Canarias, The Johns Hopkins University,
Kavli Institute for the Physics and Mathematics of
the Universe (IPMU) / University of Tokyo, Lawrence
Berkeley National Laboratory, Leibniz Institut f{\"u}r Astrophysik Potsdam (AIP), Max-Planck-Institut f{\"u}r Astronomie (MPIA Heidelberg), Max-Planck-Institut f{\"u}r Astrophysik (MPA Garching), Max-Planck-Institut f{\"u}r Extraterrestrische Physik (MPE), National Astronomical
Observatories of China, New Mexico State University,
New York University, University of Notre Dame, Observat{\'o}rio Nacional / MCTI, The Ohio State University,
Pennsylvania State University, Shanghai Astronomical
Observatory, United Kingdom Participation Group, Universidad Nacional
Aut{\'o}noma de M{\'e}xico, University of Arizona, University of Colorado
Boulder, University of Oxford, University of Portsmouth, University of
Utah, University of Virginia, University of Washington, University of Wisconsin, Vanderbilt University, and Yale University.

\clearpage
\newpage
\bibliography{zinn_arxiv_2}
\label{lastpage}
\end{document}